\documentclass[prb,10pt,aps,twocolumn,showpacs,floatfix,amssymb,amsmath]{revtex4-1}
\usepackage{rotating}

\newcommand{\beqn}{\begin{eqnarray}}
\newcommand{\eeqn}{\end{eqnarray}}
\newcommand{\eq}[1]{(\ref{#1})}

\newcommand{\bs}{\boldsymbol}
\newcommand{\Z}{{\mathbb{Z}}}

\renewcommand{\matrix}[4]{\left(\begin{array}{cc} #1 & #2 \\ #3 & #4\end{array}\right)}
\renewcommand{\vector}[2]{\left(\begin{array}{c} #1 \\ #2 \end{array}\right)}
\def\bbbone{{\mathchoice {\rm 1\mskip-4mu l} {\rm 1\mskip-4mu l} {\rm 1\mskip-4.5mu l} {\rm 1\mskip-5mu l}}}

\begin{document}

\title{On magnetic-field-induced dissipationless electric current in nanowires}
\author{M. N. Chernodub}\email{On leave from ITEP, Moscow, Russia.}
\affiliation{CNRS, Laboratoire de Math\'ematiques et Physique Th\'eorique, Universit\'e Fran\c{c}ois-Rabelais Tours,\\ F\'ed\'eration Denis Poisson, Parc de Grandmont, 37200 Tours, France}
\affiliation{Department of Physics and Astronomy, University of Gent, Krijgslaan 281, S9, B-9000 Gent, Belgium}

\begin{abstract}
We propose a general design of a metallic double-nanowire structure which may support an equilibrium dissipationless electric current in the presence of magnetic field. 
The structure consists of a compact wire element of a specific shape, which is periodically extended in one spatial dimension. Topologically, each wire element is equivalent to a ring, which supports a dissipationless current in the presence of magnetic flux similarly to the persistent electric current in a normal metal nanoring. Geometrically, each wire element breaks spatial inversion symmetry so that the equilibrium electric current through the device becomes nonzero. We also argue that the same effect should exist in long planar chiral nanoribbons subjected to external magnetic field. 
\end{abstract}


\date{May 15, 2013}

\maketitle

\section{Introduction}

A dissipationless flow of an electric current is a very desired feature in technological applications. The absence of dissipation of the electric current is usually associated with superconductivity. In a superconductor, the electric current ${\boldsymbol J}$ can be ballistically accelerated by an applied electric field ${\boldsymbol E}$ according to the London equation~\cite{ref:Tinkham}:
\beqn
\frac{\partial {\boldsymbol J}}{\partial t} = \mu^2 {\boldsymbol E}\,,
\label{eq:London}
\eeqn
where $\mu$ is a parameter related to a superconducting condensate. The electrical resistivity -- which is a convenient measure of dissipation -- is exactly zero in a superconductor. For example, in a thick superconducting ring, the accelerated electric current will continue to circulate forever, even after the applied electric field is removed.

However, there is also another, much less known, dissipationless transport law for the electric current:
\beqn
{\boldsymbol J} = {\tilde \sigma} {\boldsymbol B}\,.
\label{eq:CME}
\eeqn
Here, the electric current ${\boldsymbol J}$ is proportional to the magnitude of the magnetic field ${\boldsymbol B}$, and the proportionality coefficient $\tilde \sigma$ plays a role of conductivity. The law~\eq{eq:CME} is similar to the standard Ohm law ${\boldsymbol J} = {\sigma} {\boldsymbol E}$, in which the electric field ${\boldsymbol E}$ is substituted by the magnetic field~${\boldsymbol B}$. 

The dissipationless nature of the transport law~\eq{eq:CME} follows from a very simple fact that this equation is invariant under time inversion, $t \to -t$ [the very same property is shared by the London law~\eq{eq:London} in superconductors]. The time reversibility of Eqs.~\eq{eq:London} and \eq{eq:CME} implies that in the corresponding systems there is no entropy production associated with the electric current $\bs J$ so that this current is a dissipationless quantity in thermodynamic equilibrium.

Another feature of Eq.~\eq{eq:CME} is that this law is not invariant under inversion of the spatial coordinates, ${\boldsymbol x} \to - {\boldsymbol x}$. Thus, the dissipationless law~\eq{eq:CME} may only be realized in quite unusual, parity-odd systems which break spatial inversion symmetry. More precisely, in such systems, the spectrum of electric charge carriers should not be invariant under spatial inversion. 

We would like to mention two examples of systems where the unusual dissipationless transport law~\eq{eq:CME} may be realized. The first example comes from high energy physics, where the law~\eq{eq:CME} is known as the Chiral Magnetic Effect\cite{ref:CME}. This effect may be take place in a high energy collision of heavy ions, which creates an expanding, very hot fireball of quark-gluon plasma. Due to topological phase transitions, induced by the strong fundamental interactions, the quark-gluon plasma may contain different numbers of left- and right-handed quarks. Thus, the plasma becomes a parity-odd system. In noncentral heavy-ion collisions the quark-gluon plasma should be subjected to quite strong magnetic field. According to Eq.~\eq{eq:CME}, the parity-odd system of quarks should generate an electric current along the direction of the magnetic field. This current leaves a trace in a form of specific, charge dependent azimuthal correlations of the collision products, which were indeed observed experimentally at heavy-ion colliders\cite{ref:CME:experiment}.

In addition, the dissipationless transport law~\eq{eq:CME} is suggested to be realized in Weyl semi-metals, which admit spatial-parity-odd state(s) of charge carriers\cite{ref:Weyl}. 

\vskip 5mm
\begin{figure}[!thb]
\begin{center}
\includegraphics[scale=0.1,clip=false]{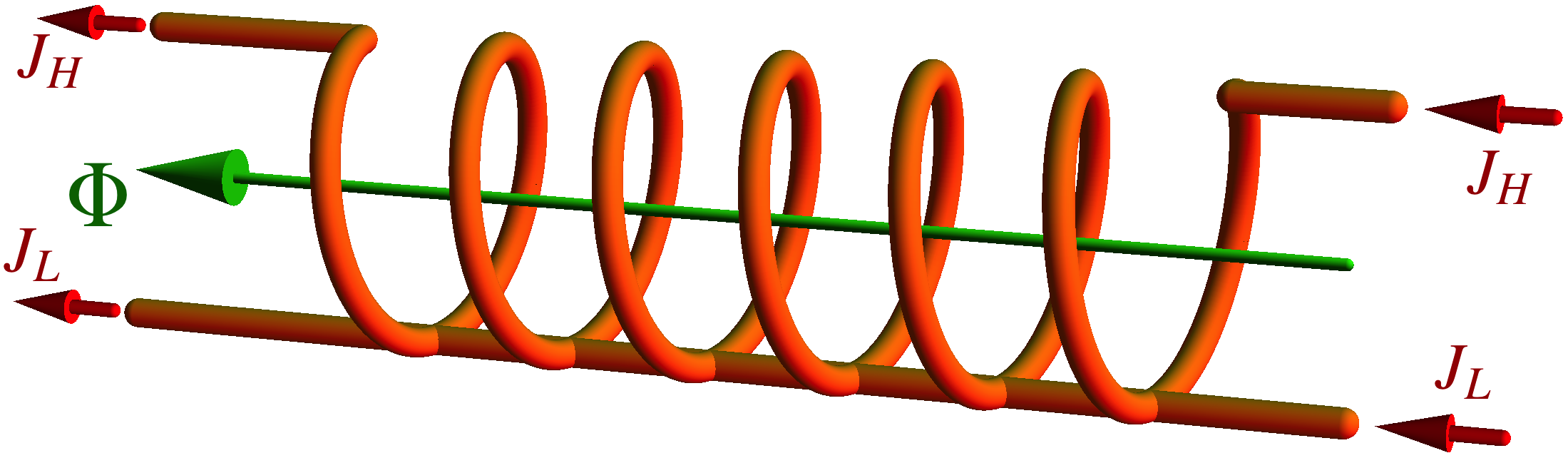} 
\end{center}
\caption{The magnetic flux $\Phi \propto B$ in this double-wire structure should generate the persistent (dissipationless) electric current $J = J_H + J_L$ even if the wires have a finite electric resistance. In a weak applied magnetic field $B$ the transport law should be given by Eq.~\eq{eq:CME}.}
\label{fig:device}
\end{figure}

In this paper we suggest that the dissipationless transport of electric current~\eq{eq:CME} may also be realized in a simple, spatially-extended device made of usual (resistive!) metal nanowires; see Fig.~\ref{fig:device}.  We show that in a background of magnetic flux $\Phi$, this device should possess (an infinite tower of) spatially odd parity states of the electric charge carriers. As a result, the electric current should flow along the magnetic field without dissipation, similarly to the persistent electric current in normal metal nanorings.

The proposed device, shown in Fig.~\ref{fig:device}, is a periodic structure made of a helix wire and a straight wire. We show in this paper that the magnetic flux $\Phi$ induces dissipationless electric current in each elementary circuit of this device, while the periodic nature of the structure makes the spectrum of the whole discrete. These effects are of a topological origin, and they appear to be related to the emergence of the persistent (dissipationless) current in normal metal nanorings~\cite{ref:Yang,ref:Kulik,ref:Buttiker}. Due to geometrical reasons, the sum of the electric currents $J_H$ and $J_L$ flowing, respectively, along the helix and linear parts of the device, is nonzero. Consequently, the total electric current $J = J_H + J_L$ passes through the whole system without dissipation. At low values of the applied magnetic field, the electric current in our system, Fig.~\ref{fig:device}, is proportional to the strength of the magnetic field~\eq{eq:CME}. 

The structure of this paper is as follows. In Sec.~\ref{sec:ring} we briefly review basic features of a dissipationless persistent current in resistive metallic nanorings in applied magnetic field. We stress the importance of both discontinuities in the energy spectrum and the parity-odd nature of the energy--current spectrum in order for the system to support the persistent electric current.

In Sec.~\ref{sec:elementary} we demonstrate that in a background of magnetic field, the short elementary device -- consisting of a single helix arc whose ends are connected together by a straight nanowire -- possesses a tower of odd--parity conducting states including the ground state. The corresponding single-level currents, passing through the device, are identified. The energy spectrum of this elementary device is, however, continuous, which does not allow for this short structure to support a persistent dissipationless electric current.

In Sec.~\ref{sec:long} we consider a long periodic structure made of an array of elementary devices, Fig.~\ref{fig:device}. We demonstrate that the energy spectrum of this long structure is a parity-odd quantity similarly to the case of the short elementary device. In addition, the spectrum of the long device becomes discontinuous (and, in fact, discrete) due to its periodic nature. These two important features, the parity-oddness and discreteness of the spectrum of the infinitely long double nanowire structure (Fig.~\ref{fig:device}) are identical to those of the electron spectrum in an ideal nanoring to be overviewed in Sec.~\ref{sec:ring}. Thus, we may expect that the proposed device should support the persistent dissipationless current in accordance with Eq.~\eq{eq:CME}. Notice that the presence of the disorder and electron interactions in real, non-ideal nanorings does not destroy the persistent current completely so that the dissipationless motion of electrons survives even in the real metal rings with the finite resistivity. The thermal fluctuations suppress exponentially the magnitude of the current thus, given our experience with the nanorings, the suggested effect should be accessible experimentally in a reasonable temperature interval. 

In Sec.~\ref{sec:twodim} we also briefly discuss a two-dimensional (flat) analogue of the our double-wire system, Fig.~\ref{fig:device}, which turns out to be a chiral (nano)ribbon visualized in Fig.~\ref{fig:twodim}. In this case the analogue of the dissipationless law~\eq{eq:CME} is given by Eq.~\eq{eq:alternative:CME}.

We conclude in Sec.~\ref{sec:conclusions} that the proposed systems, shown in Figs.~\ref{fig:device} and \ref{fig:twodim}, may support the dissipationless current, Eqs.~\eq{eq:CME} and \eq{eq:alternative:CME}, respectively, even in real resistive metal nanorings at finite temperature.

\section{Persistent electric currents in normal metal rings}
\label{sec:ring}

\subsection{Energy spectrum and single-level currents}

In this section we briefly overview the appearance of the persistent electric current in a normal, nonsuperconducting metallic ring (see Refs.~[\onlinecite{ref:reviews:nanorings},\onlinecite{ref:Thesis}] and references therein). 

\begin{figure}[!thb]
\begin{center}
\includegraphics[scale=0.1,clip=false]{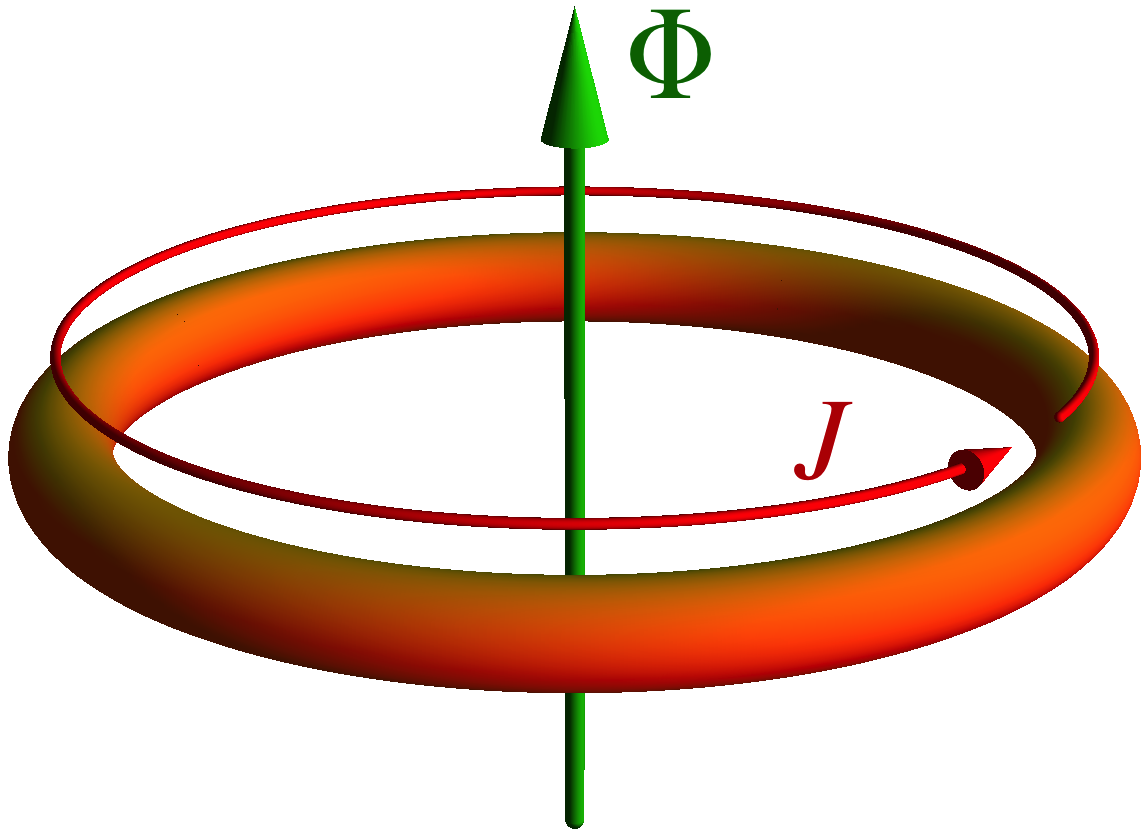} 
\end{center}
\vskip -5mm
\caption{Metallic ring threaded by a constant magnetic flux $\Phi$ carries persistent electric current $J$.}
\label{fig:quantumring}
\end{figure}

We consider an ideal one-dimensional ring of the radius $R$ which is pierced by the magnetic flux~$\Phi$; see Fig.~\ref{fig:quantumring}. The electrons in the ring are assumed to be noninteracting. The electromagnetic vector potential can conveniently be chosen in the form $A_0 = A_3 = 0$ and $\boldsymbol{A} = 2 \pi \Phi {\hat{\boldsymbol \varphi}}/L^2_R$ where $L_R = 2 \pi R$ is the total circumference of the ring and ${\hat{\boldsymbol \varphi}}$ is the unit vector in the azimuthal $\varphi$ direction (the only spatial degree of freedom of the electron is the azimuthal angle $\varphi$). 

The eigenfunctions $\Psi_n$ and the corresponding energies $E_n$ of a single electron are determined by the time-independent Schr\"odinger equation:
\beqn
- \frac{\hbar^2}{2 m} \left(\frac{\partial}{\partial \xi} + \frac{2 \pi i}{L_R} \frac{\!\!\Phi}{\Phi_0} \right)^2 \Psi_n = E_n \Psi_n\,,
\label{eq:ring:equation}
\eeqn
where $\Phi_0 = h/e$ is the flux quantum and $\xi = R \varphi $ is the convenient spatial variable which varies from $0$ to~$L_R$. The fermionic nature of electrons leads to the double degeneracy of the energy levels while other spin-related effects (such as small Zeeman splitting of the levels) are usually ignored in this simplified but, nevertheless, working approach.

Due to the periodicity of the wave functions, 
\beqn
\Psi_n(L) = \Psi(0)\,,
\label{eq:ring:periodicity}
\eeqn
the Schr\"odinger equation~\eq{eq:ring:equation} has the discrete spectrum:
\beqn
\Psi_n(\xi) & = & \frac{1}{\sqrt{L_R}} \exp\left( 2 \pi i n \frac{\xi}{L_R} \right) \,,  
\label{eq:psi:n} \\ 
E_n & = & \frac{h^2}{2 m L^2_R} \left( n + \frac{\!\!\Phi}{\Phi_0} \right)^2 
\equiv \frac{\hbar^2}{2 m R^2} \left( n + \frac{\!\!\Phi}{\Phi_0} \right)^2 \!,
\label{eq:epsilon:n}
\label{eq:E:n}
\eeqn
characterized by the integer $n \in \Z$.

\begin{figure}[!thb]
\begin{center}
\includegraphics[scale=0.35,clip=false]{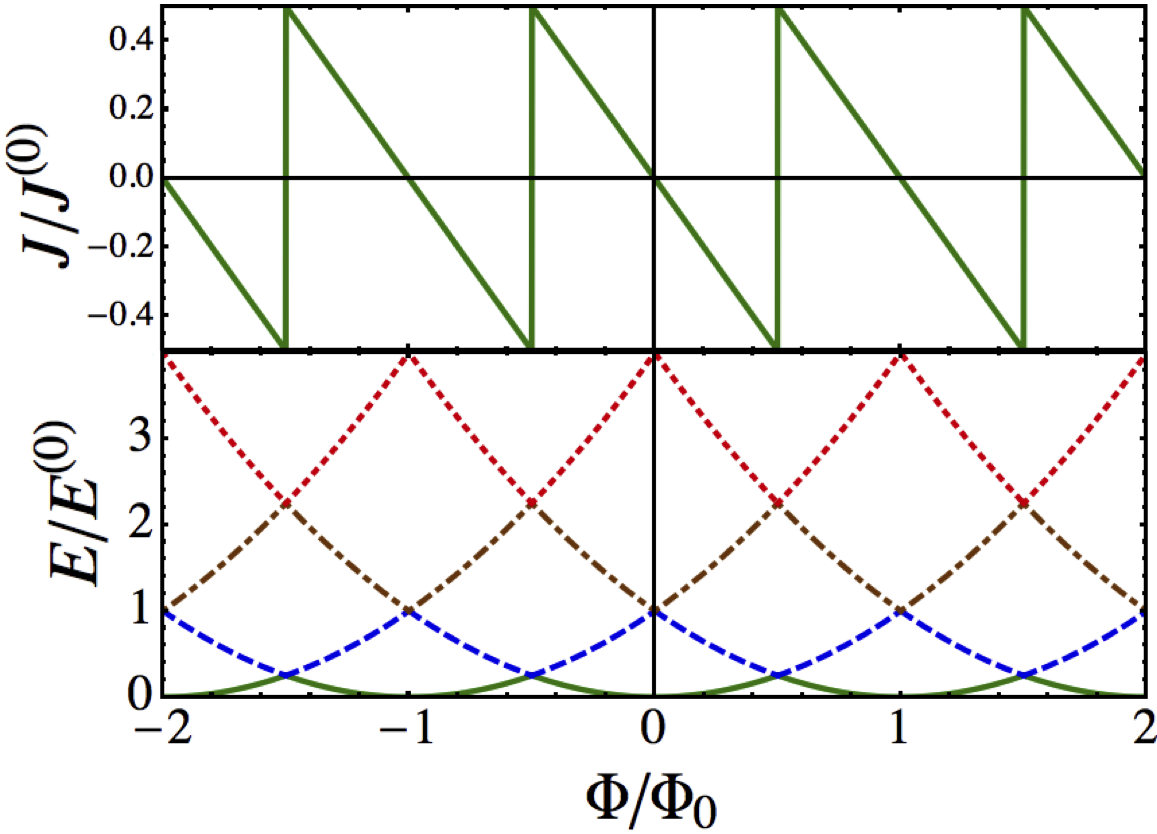} 
\end{center}
\caption{(top) Single-level electric current [shown in units of $J^{(0)} = e h/(m L^2_R)$] in the ground state and (bottom) first four levels of the energy spectrum [in units of $E^{(0)} = h^2/(2 m L^2_R)$] versus total magnetic flux $\Phi$ (in units of the elementary flux $\Phi_0 = h/e$) in a perfect one-dimensional ring.}
\label{fig:current:energy}
\end{figure}

The energy spectrum~\eq{eq:epsilon:n} of the perfect one-di\-men\-sio\-nal ring exhibits a periodic dependence on the Aharonov-Bohm flux $\Phi$; the first four energy levels are illustrated in the bottom panel of Fig.~\ref{fig:current:energy}. 

The single-level electric current $J_n$ carried by an electron in $n$th eigenstate~\eq{eq:psi:n} is given by the following formula:
\beqn
J_n \equiv \langle n | \hat J | n \rangle = - \frac{\partial \varepsilon_n}{\partial \Phi} 
= - \frac{e h}{m L^2_R} \left( n + \frac{\!\!\Phi}{\Phi_0} \right)\,. 
\label{eq:i:n}
\eeqn
We do not consider the trivial doubling effect due to the spin degeneracy [in this case the current in Eq.~\eq{eq:i:n} should be multiplied by a factor of 2].

\subsection{Persistent electric current}

What are the signatures that the real quantum ring may support the persistent electric current?

Firstly, it is important to notice that the ground-state electric current in the ring is nonzero for general values of the magnetic flux~$\Phi$; see the top panel of Fig.~\ref{fig:current:energy} for an illustration. The presence of the electric current in the ground state of this idealized system provides us with a signal that a real system (given by a finite-width resistive wire at finite temperature with effects of both disorder and interactions included) should support a persistent electric current as well~\cite{ref:Buttiker}. 

\vskip 5mm
\begin{figure}[!thb]
\begin{center}
\includegraphics[scale=0.12,clip=false]{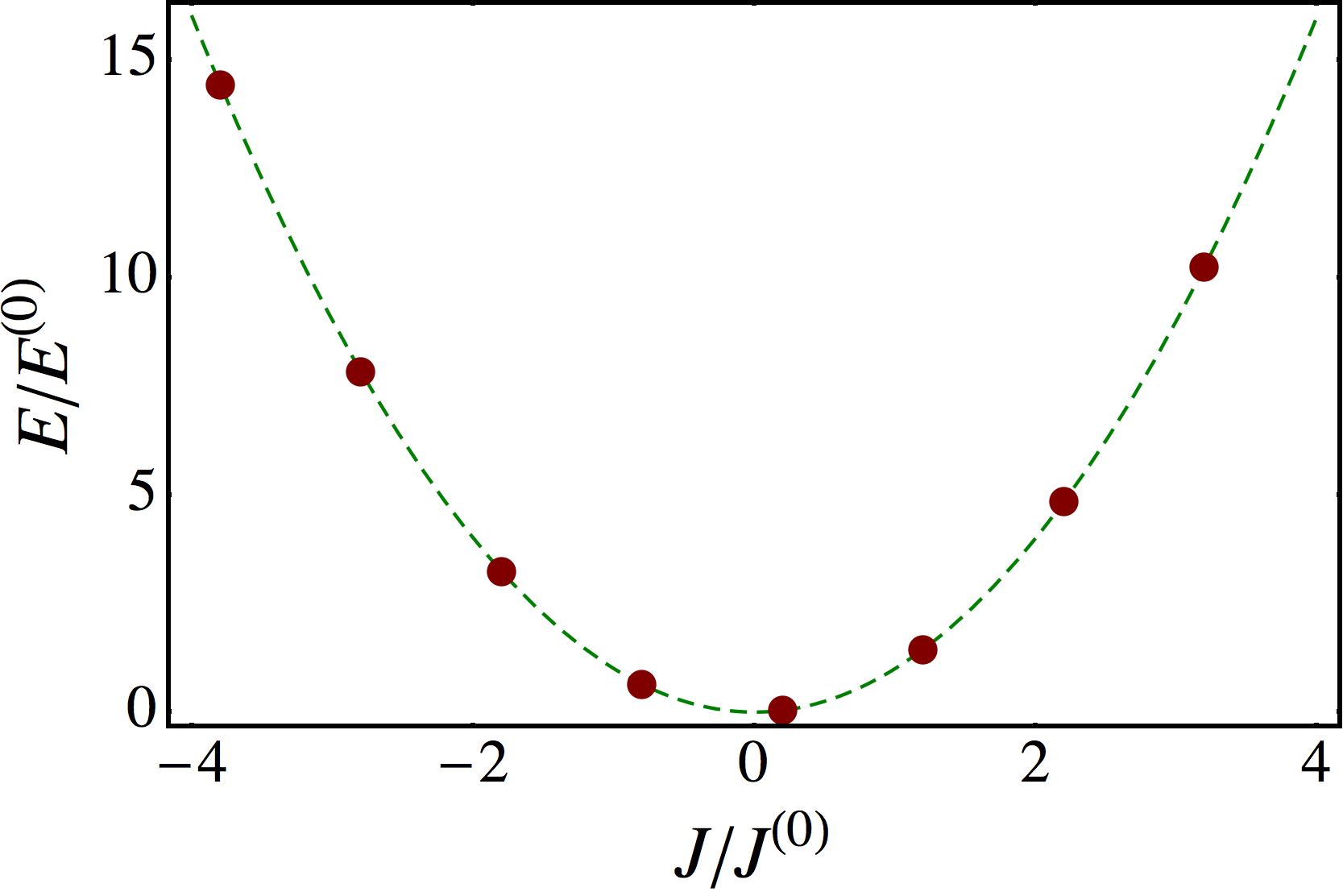} 
\end{center}
\caption{Energy of the elementary excitations~\eq{eq:E:n} vs. the corresponding single-level currents~\eq{eq:i:n} in the ideal one-dimensional ring [units of $E^{(0)} = h^2/(2 m L^2_R)$ and $J^{(0)} = e h/(m L^2_R)$, respectively] pierced by the magnetic flux $\Phi = 0.2\, \Phi_0$.}
\label{fig:current:vs:energy}
\end{figure}

Secondly, higher energy levels have an asymmetric structure. In Fig.~\ref{fig:current:vs:energy} we show a typical behavior of the energy levels of the system~\eq{eq:E:n} as a function of the current~\eq{eq:i:n}. Although the energy levels follow a symmetric parabola, $E \propto J^2$, the levels are in fact not symmetric under the inversion of electric current ($J_n \to - J_n$) provided that the ratio $\Phi/(2 \Phi_0)$ is not equal to an integer number. 

Thirdly, in order to support the persistent current, the energy spectrum of the system should be either a piecewise discontinuous function or a discrete function of momenta because for a general continuous spectrum the net contribution to the equilibrium electric current will always be zero\cite{footnote0}.

In a real metallic ring at zero temperature, the electrons occupy all energy levels up to the Fermi energy $E_F$. Each electron possesses its own single-level electric current which circulates inside the ring. In general, the total electric current $J \equiv \langle J \rangle$, given by the sum of the individual single-level contributions, is nonzero. In the sum, the individual single-level currents almost cancel each other, and the typical electric current, $J \sim e v_F / L_R$, is determined by the Fermi velocity $v_F$. Thus, the effect is given by the dynamics of a few electrons which are residing close to the Fermi surface\cite{ref:Cheung}.

At finite temperature, the electron's distribution is given by the Fermi-Dirac statistics so that the higher energy levels with $E > E_F$ are excited and they also contribute to the total electric current. Both the magnitude and the sign of the single-level electric current evolve rapidly as its energy increases, so that temperature increase gives rise to the sign-alternating contributions to the total current $\langle J \rangle$. As a result, theoretically, the average electric current should drop exponentially as the function of temperature, both in the absence\cite{ref:Kulik} or presence of diffusion caused by impurities\cite{ref:Riedel}. 

Nevertheless, neither the temperature effects, nor electron interactions, nor the presence of diffusion, nor finite-width effects of real nanowires do not completely destroy the induced electric current.  Thus, in general, an interacting many-electron system in the normal metallic ring should possess a nonzero persistent average electric current, which circulates without dissipation. Due to the absence of the dissipation, the current should, in general, be circulating forever, hence its name ``persistent current''.

The persistent current is a few-electron phenomenon which is generally affected by (both quantum and thermal) fluctuations. In a given ring, the average current $\langle J \rangle$ is a finite quantity, which is usually much smaller than the typical current $\langle J^2 \rangle^{1/2}$. Due to the strong dependence of the persistent current on the values of the magnetic flux and occupation number, both the magnitude and the sign of the current vary with the number of the electrons in the sample (and nominally identical rings may have have unequal persistent currents in the same environment).  A typical magnitude of the measured persistent current in the experimentally accessible metallic rings varies, in order of magnitude, from $1\,{\mathrm{pA}}$  ($10^{-12}\,{\mathrm{A}}$) to $1\,{\mathrm{nA}} $ ($10^{-9}\,{\mathrm{A}}$) in a range of temperatures up to 3 K. Excellent experimental results -- which confirm the existence of the  persistent currents in metallic nanorings -- were obtained recently in Refs.~[\onlinecite{ref:Science},\onlinecite{ref:PRL}] which follow the extensive set of earlier experimental results\cite{ref:earlier:experiments}.

The important and obvious property of the persistent current in the ring is that it circulates in (and confined to) a plane which is transverse with respect to the direction of the magnetic field. In the next sections we consider the simple geometrical generalization of the metallic ring, Fig.~\ref{fig:device}, which may allow for the electric current to flow without dissipation {\emph{along}} the direction of the magnetic field. 

\section{Energy levels in a short wire}
\label{sec:elementary}

\begin{figure}[!thb]
\begin{center}
\includegraphics[scale=0.05,clip=false]{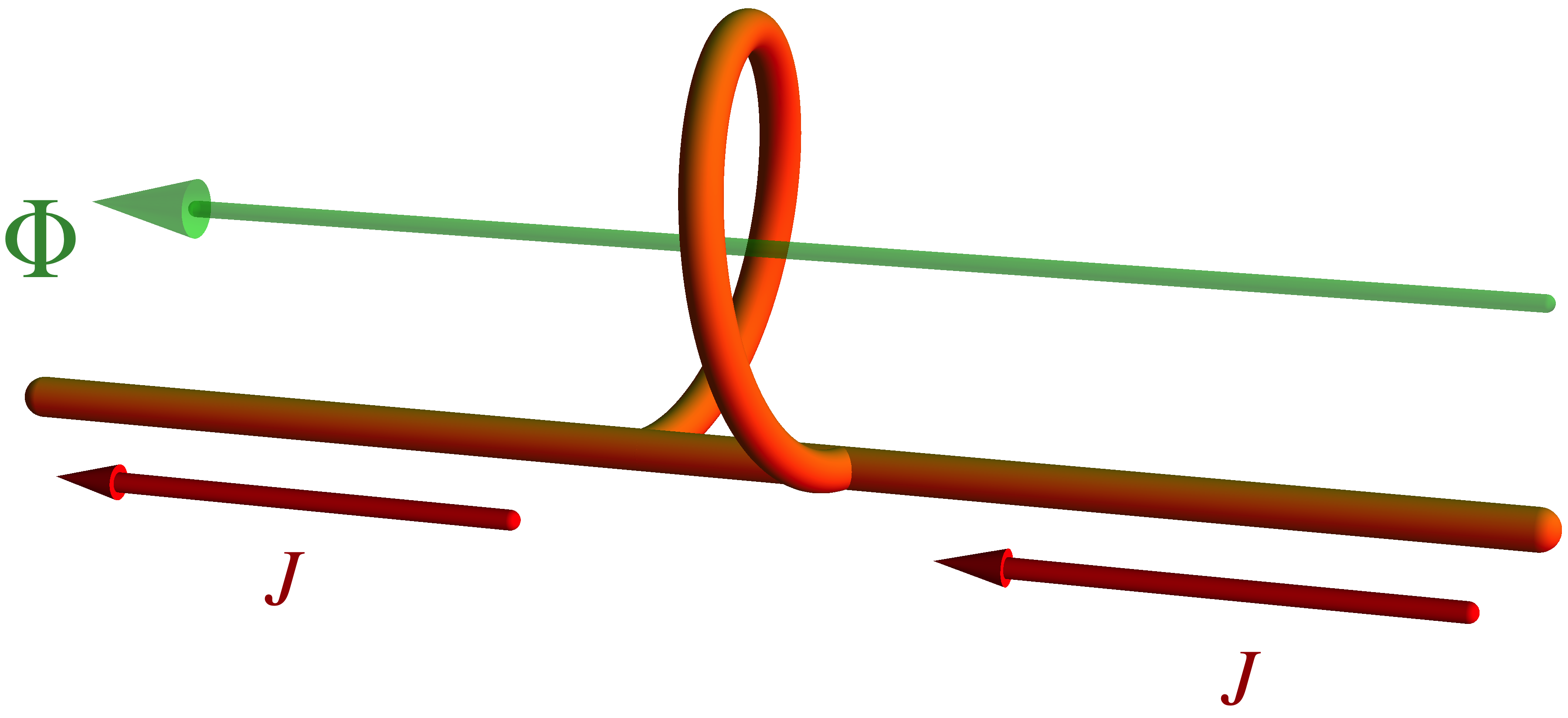} 
\end{center}
\vskip -5mm
\caption{One-component device in external magnetic flux $\Phi$.}
\label{fig:experiment}
\end{figure}

In this section we discuss the spectrum of an elementary device which is made of a single arc of helix and a straight wire, as it is shown in Fig.~\ref{fig:experiment}. This simple device is a part of the more general construction  shown in Fig.~\ref{fig:device}. We will show that in the presence of magnetic flux $\Phi$, the energy spectrum of this device $E(J)$ is  not symmetric with respect to the inversion of the electric current $J \to - J$ propagating through the device. 

\begin{figure}[!thb]
\begin{center}
\includegraphics[scale=0.6,clip=false]{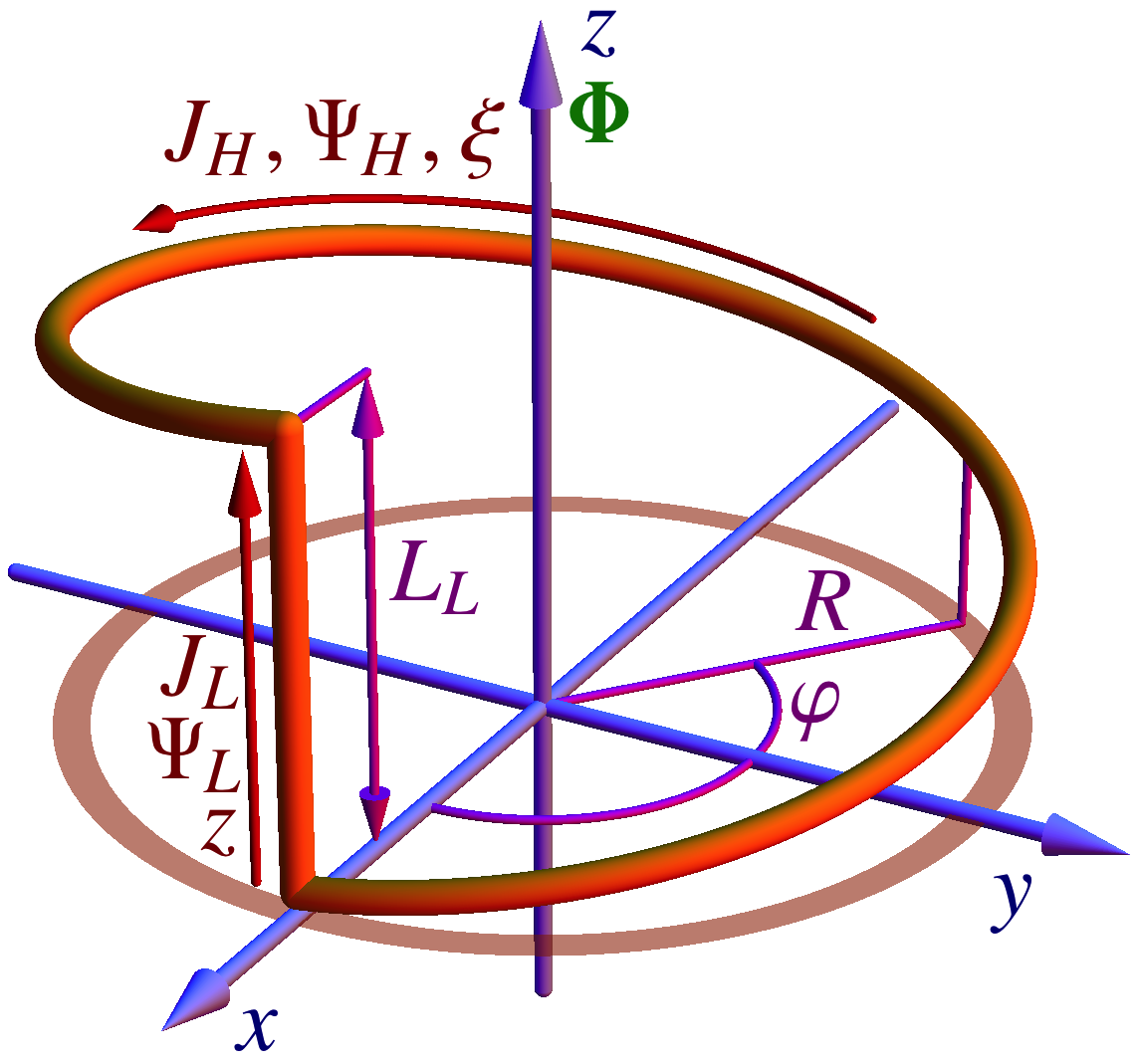} 
\end{center}
\vskip -5mm
\caption{Elementary device: coordinates and quantities.}
\label{fig:coordinates}
\end{figure}

Our coordinate system and other notations are illustrated in Fig.~\ref{fig:coordinates}. The helical part of the device is described by the following parametric formula: 
\beqn
x = R \cos \varphi\,, \quad 
y = R \sin \varphi\,, \quad 
z = \frac{L_L \;\! \varphi}{2\pi}\,, \quad 
\eeqn
where $R$ is the radius of helix, $L_L$ is height of the elementary arc of the helix and $\varphi \in [0,2\pi]$ is the angle which parameterizes the arc. The length of the elementary helix arc~is 
\beqn
L_H = \sqrt{L_L^2 + L_R^2}\,, \qquad L_R = 2 \pi R\,.
\label{eq:LH:L}
\eeqn

It is convenient to introduce the natural spatial coordinate for the helix arc, $\xi \in [0,L_H]$, and the coordinate for the straight element, $z \in [0,L_L]$. 

The wave function of the electron in this device is a vector function:
\beqn
\Psi_n(\xi,z) = \vector{\Psi_{H,n}(\xi)}{\Psi_{L,n}(z)} \,.
\label{eq:Psi:n:vector}
\eeqn
The upper (lower) component of this vector wavefunction describes the wavefunction of the electron in the helix arc (line) part of the device.

As in the case of the metallic ring ring, we consider noninteracting electrons, ignore disorder and spin effects. Also, for the sake of simplicity, we assume that the straight part  (L) of our device and its helix part (H) are made of the same material so that the effective electron masses in both parts the same: $m = m^*_L = m^*_H$.

Thus, the electrons in our elementary device are described by the Schr\"odinger equation,
\beqn
{\hat H}(\xi,z)  \Psi_n(\xi,z)  = E_n \Psi_n(\xi,z) \,,
\label{eq:Schroedinger}
\eeqn
where the Hamiltonian has a form of a diagonal matrix:
\beqn
{\hat H}(\xi,z) & = & \matrix{{\hat H}_H(\xi)}{0}{0}{{\hat H}_L(z)}\,.
\eeqn
Here the individual Hamiltonians 
\beqn
{\hat H}_H(\xi) & = & - \frac{\hbar^2}{2 m} \left[\left(\frac{\partial}{\partial \xi} + \frac{2 \pi i}{L_H} \frac{\!\!\Phi}{\Phi_0} \right)^2 - \frac{\kappa^2}{4}\right]\!, 
\label{eq:H:H}\\
{\hat H}_L(z) & = & - \frac{\hbar^2}{2 m} \frac{\partial^2}{\partial z^2}\,,
\label{eq:H:L}
\eeqn
describe the dynamics of free electrons in, respectively, the helix-arc part and the line part of the device.

The last term in brackets of the helix arc Hamiltonian~\eq{eq:H:H} is the constant term which appears due to the arc curvature~\cite{ref:daCosta}:
\beqn
\kappa = \frac{2 \pi L_R}{L_H^2}\,.
\label{eq:curvature}
\eeqn
In nanorings, the curvature term ($\kappa_{\mathrm{ring}} = 1/R$) is neglected since it leads to the same constant shift for all energy levels. However, in our case this term will play an important quantitative role. 

Similarly to the matching condition~\eq{eq:ring:periodicity} of the wavefunctions at the ring, the Sch\"odinger equation~\eq{eq:Schroedinger} should also be complemented with the matching conditions at the points where the arc part and the line part of the device intersect with each other; see Fig.~\ref{fig:coordinates}. There are two such points, $(\xi,z) = (0,0)$ and $(\xi,z) = (L_H,L_L)$, and the corresponding matching conditions are as follows:
\beqn
\Psi_H(0) = \Psi_L(0)\,, 
\qquad 
\Psi_H(L_H) = \Psi_L(L_L)\,.
\label{eq:matching}
\eeqn

In our idealized approach we treat the wires as infinitesimally thin linear objects, and we do not consider possible scattering effects at the junctions. For the same reason, we do not consider possible effects of the curvature at the junction points. A more detailed discussion on the scattering at the junctions will be presented in the next section.

\begin{figure}[!thb]
\begin{center}
\includegraphics[scale=0.3,clip=false]{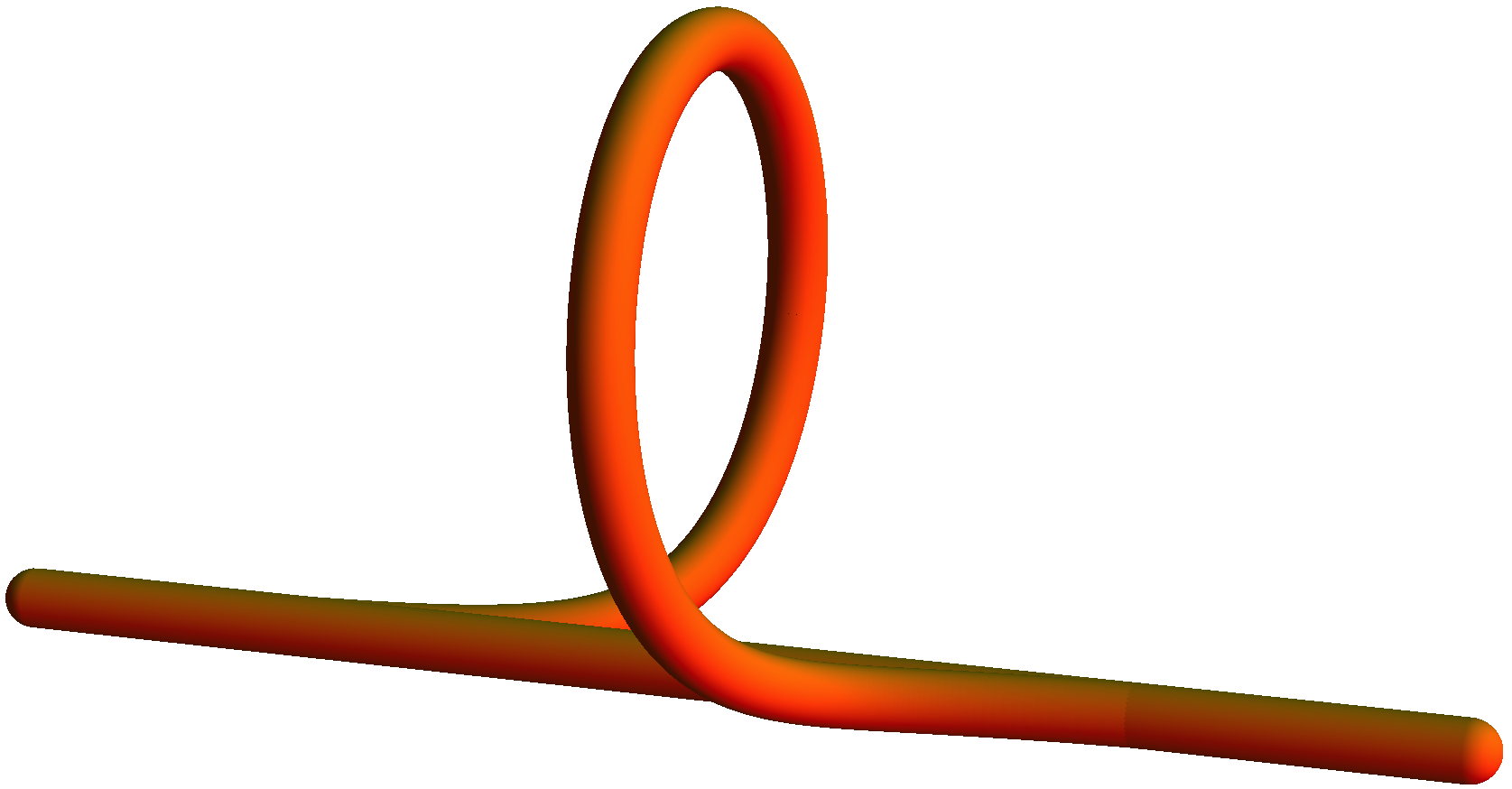} 
\end{center}
\vskip -5mm
\caption{The elementary device, Fig.~\ref{fig:experiment}, with smooth junctions.}
\label{fig:device:smooth}
\end{figure}

In realistic applications the junctions can be made smoother as it is visualized in Fig.~\ref{fig:device:smooth}. 
In this case the curvature of the arc is no more constant, $\kappa = \kappa(\xi)$, and the curvature term in Eq.~\eq{eq:H:H} should act as a potential imposed on the charge carries. This term leads to additional known curvature effects such as, for example, the appearance of curvature-induced bound states and resonances\cite{ref:resonances}. In our we rather concentrate on important topological effects of the junctions themselves.

A general solution of Eq.~\eq{eq:Schroedinger} is given by the following set of wave functions:
\beqn
\Psi_H & = & C^A_H e^{i(q_H - \gamma) \xi} + C^B_H e^{- i(q_H + \gamma) \xi}\,, \\
\Psi_L & = & C^A_L e^{i q_L z} + C^B_L e^{- i q_L z}\,,
\label{eq:anzats:elementary}
\eeqn
where $q_H$ and $q_L$ are the momenta in the helix arc and in the line wire, and $C^A_H$, $C^B_H$, $C^A_L$ and $C^B_L$ are the parameters which should be fixed by the normalization condition for the wavefunction,
\beqn
\int_0^{L_H} d \xi \, |\Psi_H(\zeta)| + \int_0^{L_L} d z \, |\Psi_L(z)| = 1 \,,
\label{eq:normalization}
\eeqn
and by the matching conditions~\eq{eq:matching}. The parameter
\beqn
\gamma = \frac{2 \pi}{L_H} \frac{\Phi}{\Phi_0}
\label{eq:gamma:Phi}
\eeqn
depends on the total flux $\Phi$ of the magnetic field which pierces the helix arc. 

The energy of the electrons in the device is given by the following expression,
\beqn
E_q = \frac{\hbar^2}{2 m} \left(q^2_H - \frac{\kappa^2}{4} \right) = \frac{\hbar^2 q^2_L}{2 m}\,,
\label{eq:energy}
\eeqn
which also sets the constraint on the momenta $q_H$ and~$q_L$.

The energy spectrum of the electronic excitations~\eq{eq:energy} in the elementary device is determined by Eqs.~\eq{eq:matching}, \eq{eq:anzats:elementary} and \eq{eq:normalization}. Since the aim of this section is to demonstrate, qualitatively, the parity-odd nature of the magnetic-field-induced longitudinal electric current, we simplify the problem by restricting ourselves to the one-wave solution with $C^B_H = 0$. This choice is justified by the fact that in the limit $L_L \to 0$ our device reduces to the closed ring (see Fig.~\ref{fig:coordinates}) and in the latter case the spectrum of the electron excitations in the ring is given by Eq.~\eq{eq:psi:n} which corresponds to our one-wave anzats with $C^B_H = 0$. For the most interesting case of the infinitely long ``multi-device'', Fig.~\ref{fig:device}, a full set of solutions, extending beyond the one-wave anzats, will be given in the next section.

The matching conditions~\eq{eq:matching} provide us with the following constraints on the remaining parameters $C^A_H$, $C^A_L$ and $C^B_L$ of our anzats~\eq{eq:anzats:elementary}:
\beqn
C^A_H & = & C^A_L + C^B_L\,, \nonumber \\
C^A_H e^{i (q_H - \gamma) L_H} & = & C^A_L e^{i q_L L_L} + C^B_L e^{- i q_L L_L}\,. \qquad
\eeqn
These equations can be solved as follows:
\beqn
C^A_L & = & R(q_L,q_H) \, C^B_L\,, \\
C^A_H & = & \left[1 + R(q_L,q_H) \right] C^B_L\,,
\eeqn
where
\beqn
R(q_L,q_H) = \frac{e^{-i q_L L_L} - e^{i(q_H - \gamma) L_H}}{e^{i(q_H - \gamma) L_H} - e^{i q_L L_L}}\,.
\label{eq:R:q}
\eeqn

The normalization condition for the wavefunction~\eq{eq:normalization}, gives us the normalization parameter $C^B_L$:
\beqn
C^B_L = 1/\sqrt{N(q_L,q_H)}\,,
\label{eq:N:B}
\eeqn
where
\beqn
& & N(q_L,q_H) = |1 + R(q_L,q_H)|^2 L_H \nonumber \\
& & \hskip 16.5mm + \left( 1 + |R(q_L,q_H)|^2 \right) L_L  \label{eq:N:q} \\
& & \hskip 3mm + \left\{{\mathrm{Im} \left[R(q_L,q_H) e^{2 i q_L L_L}\right]} - {\mathrm{Im} \left[R(q_L,q_H)\right]}\right\} q^{-1}_L\,.
\nonumber
\eeqn
We chose the parameter $C^B_L$ to be a real number since its phase corresponds to a coordinate-independent, inessential phase of the whole solution. 

The one-wave spectrum of the electrons in our elementary device is parameterized by the continuous momentum $q_H$ in the helix arc, while the momentum $q_L$ in the line element is defined via Eq.~\eq{eq:energy}.

The electric currents in the helical arc part and in the line wire part are, respectively, as follows:
\beqn
J_H & = & - \frac{e \hbar q_H}{m} \frac{|1 + R(q_L,q_H)|^2}{N(q_L,q_H)}\,, 
\label{eq:J:H:elem}\\
J_L & = & - \frac{e \hbar q_L}{m} \frac{|R(q_L,q_H)|^2 - 1}{N(q_L,q_H)}\,,
\label{eq:J:L:elem}
\eeqn
and the total current passing through the device in the direction of the magnetic field is given by the sum of currents passing through the helix arc and line wire:
\beqn
J & \equiv & J_H + J_L \,.
\label{eq:J:tot:elem}
\eeqn
Equations~\eq{eq:energy}, \eq{eq:R:q}, \eq{eq:N:q} and \eq{eq:J:tot:elem} allow us to obtain a parametric (parameterized by $q_H$) relation between the level energy~\eq{eq:energy} and the total single-level electric current~\eq{eq:J:tot:elem}.

For illustration purposes we set the radius of our helix $R$ to be equal to a typical radius of the ring of Ref.~[\onlinecite{ref:Science}], $R = 500$\,nm. One unit of the magnetic flux $\Phi(B_0) = \Phi_0$ for this value of the radius is reached at the magnetic field $B_0 = 5.3\cdot 10^{-3}\,{\mathrm{T}} \equiv 53\,{\mathrm{G}}$.
\begin{figure}[!thb]
\begin{center}
\includegraphics[scale=0.125,clip=false]{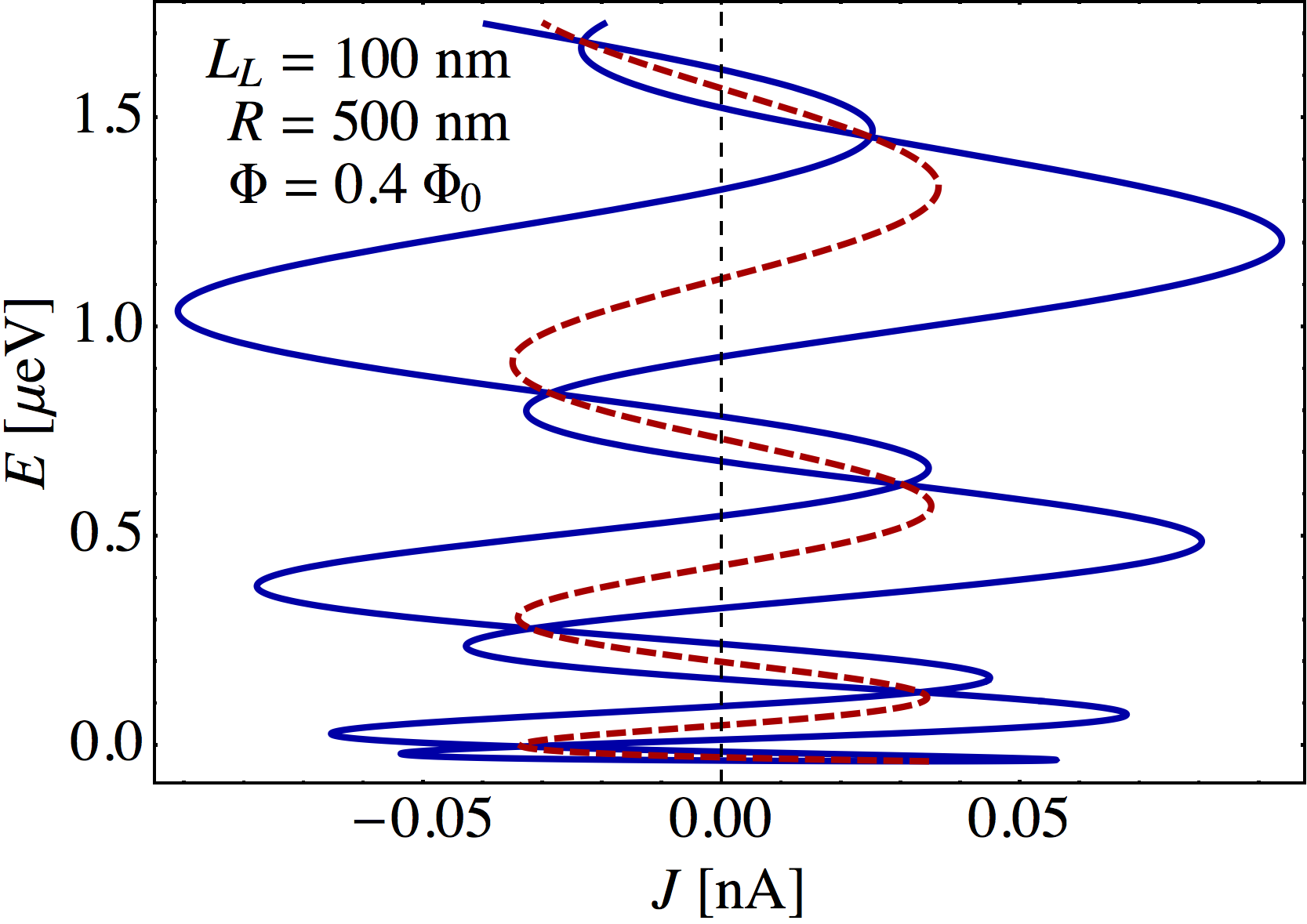} 
\end{center}
\vskip -5mm
\caption{Energy of the elementary excitations~\eq{eq:energy} vs. the corresponding total one-wave single-level current~\eq{eq:J:tot:elem} passing through the elementary device, Fig.~\ref{fig:experiment}, is shown by the blue solid line. The red dashed line represents the level's energy $E$ as the function of the averaged current ${\bar J}$ of the same energy (see the description in the text). The current is induced by the magnetic flux $\Phi = 0.4 \, \Phi_0$. The device dimensions are $R = 500\,\mbox{nm}$ and $L_L = 100\,\mbox{nm}$.}
\label{fig:spectrum:full}
\end{figure}
\begin{figure}[!thb]
\begin{center}
\includegraphics[scale=0.125,clip=false]{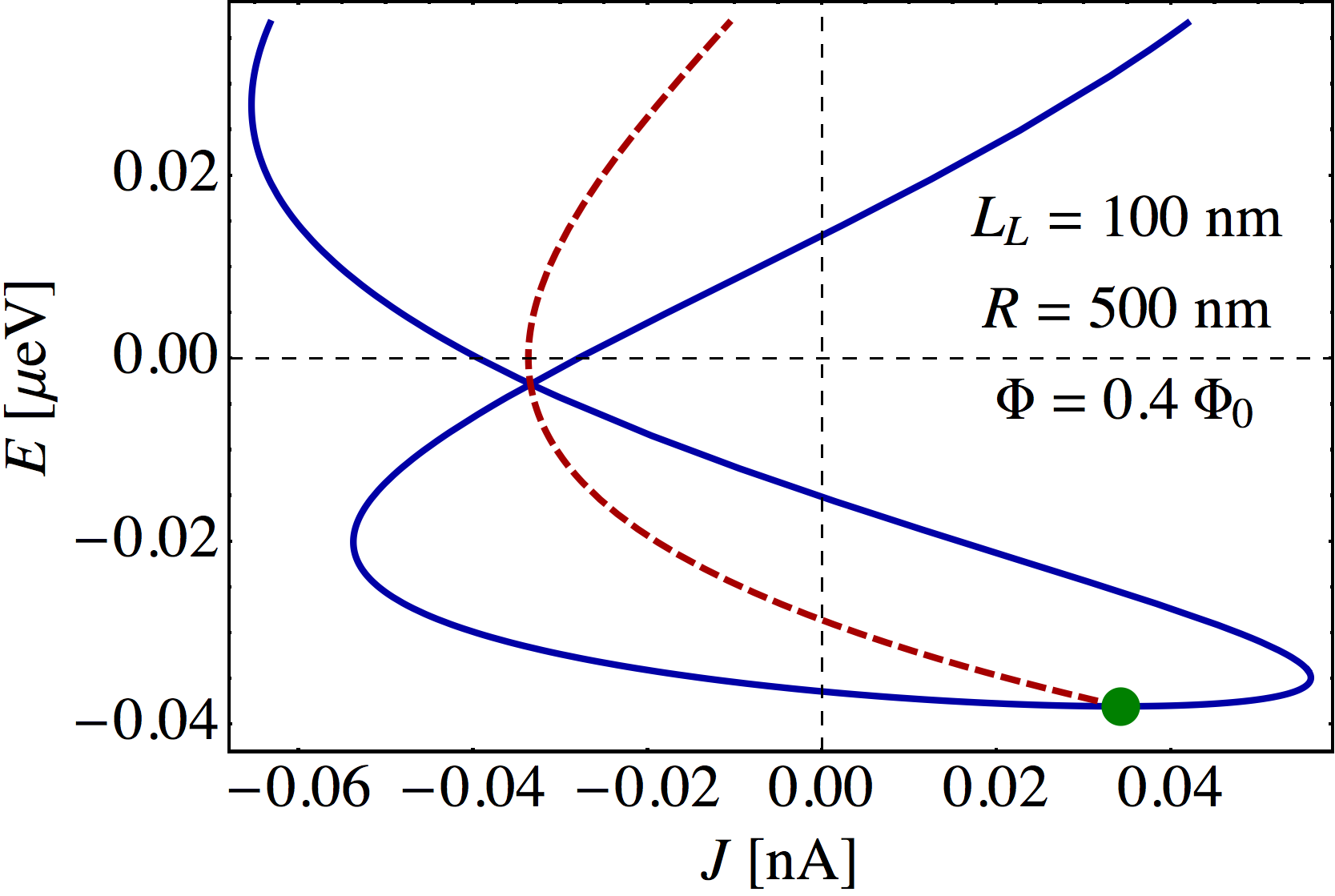} 
\end{center}
\vskip -5mm
\caption{The lower part of the spectrum of Fig.~\ref{fig:spectrum:full}. The green dot corresponds to the single-level persistent current in the ground state~\eq{eq:current0}. It is worth noticing the continuous nature of the bound state spectrum with $E < 0$.}
\label{fig:spectrum:low}
\end{figure}

As an example, in Fig.~\ref{fig:spectrum:full} and Fig.~\ref{fig:spectrum:low} we show the continuous one-wave energy spectrum for the elementary device with the size of the linear element $L_L = 100\,\mbox{nm}$ at the magnetic flux $\Phi = 0.4 \, \Phi_0$. Notice that the spectrum (shown by the solid blue line) is infinitely degenerate because the same value of the total electric $J$ current is realized by infinitely many energy states $E$.  Moreover, a fixed value of energy $E$ corresponds, in general, to two values, $J_+$ and $J_-$,  of the total electric current, with $E = E(J_+) = E(J_-)$. The energy $E$ as the function of the average current, ${\bar J} = J_+ + J_-$ is shown in Fig.~\ref{fig:spectrum:full} and Fig.~\ref{fig:spectrum:low} by the dashed red curve.

One can clearly see that the energy spectrum is not invariant under the reflection of the total current~\eq{eq:J:tot:elem} which passes throughout the device, $J \to - J$. This important property of the elementary excitations in our elementary device is very similar to the property of the energy spectrum in nanorings, Fig.~\ref{fig:current:vs:energy}. It is the non-invariance of the energy spectrum which is responsible for the persistent electric current in resistive, normal metal nanorings. 

In the limit of low energies and momenta, $q_H \to 0$, we get the following expression for the ground state current~\eq{eq:J:tot:elem}:
\beqn
& & J_0 \equiv J(q_H \to 0) = \kappa ^2 \sinh  \left(\frac{\kappa   L_L}{2}\right) \sin \left(\frac{2 \pi  \Phi}{\Phi_0} \right) 
\label{eq:current0} \\
& & \cdot \biggl[\kappa   L_H \cosh (\kappa L_L)+2 \sinh (\kappa   L_L)  - \kappa (L_H + 2  L_L)
\nonumber \\
& & + 4 \left(\frac{\kappa   L_L}{2} \cosh\frac{\kappa   L_L}{2} - \sinh \frac{\kappa   L_L}{2}\right) \cos \frac{2 \pi  \Phi}{\Phi_0}\biggr]^{-1}\!,
\nonumber 
\eeqn
where the curvature $\kappa$ is given in Eq.~\eq{eq:curvature} and we have also expressed the quantity $\gamma$ in terms of the magnetic flux $\Phi$ according to Eq.~\eq{eq:gamma:Phi}. The ground-state current for the device with $R=500\,\mbox{nm}$ and $L_L = 100\,\mbox{nm}$ at the fixed flux $\Phi = 0.4 \Phi_0$ is shown in Fig.~\ref{fig:spectrum:low} by a green circle.

\begin{figure}[!thb]
\begin{center}
\includegraphics[scale=0.125,clip=false]{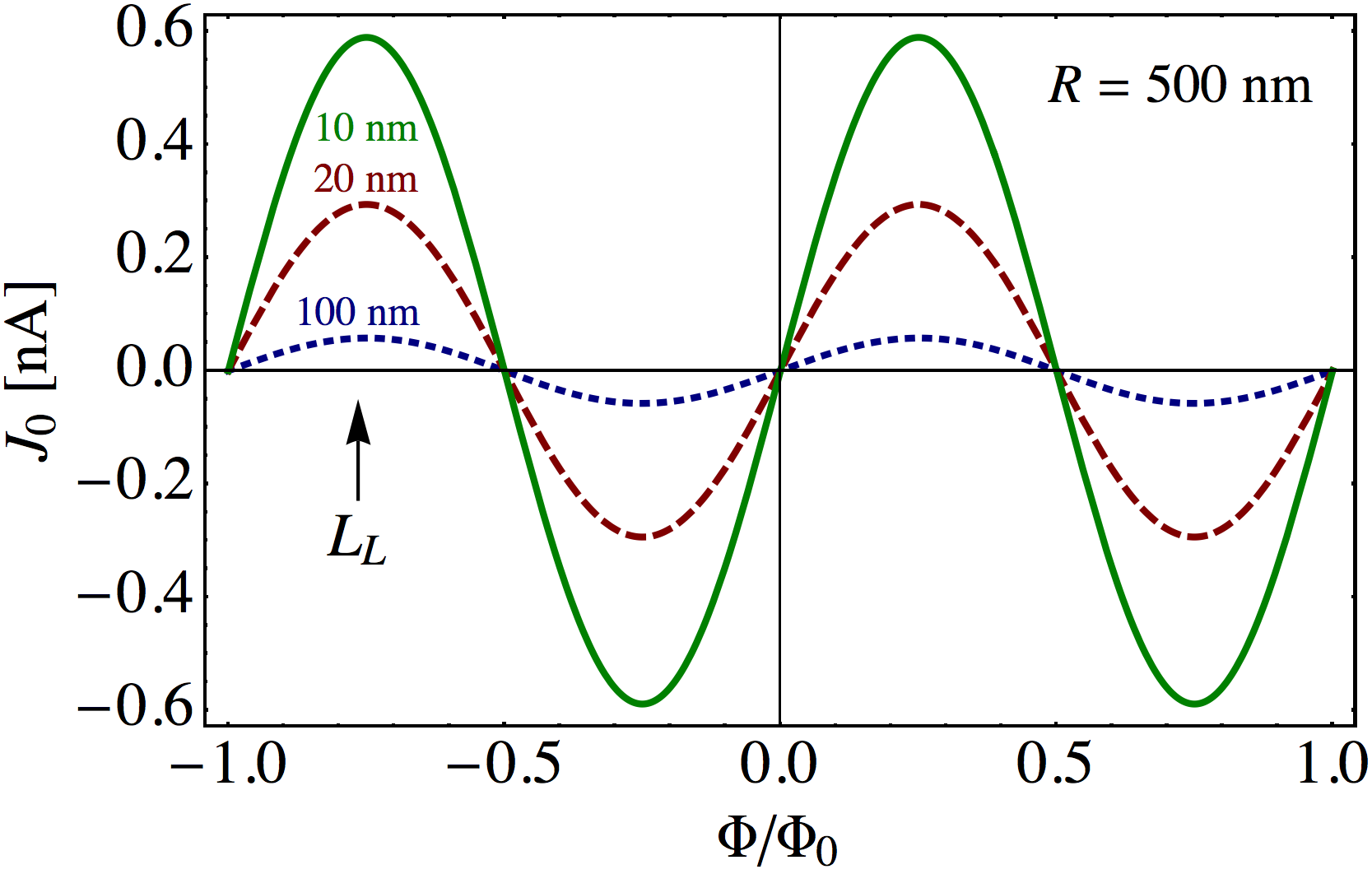} 
\end{center}
\vskip -5mm
\caption{Ground state current for the one-wave electron excitations in the elementary device as a function of the magnetic flux $\Phi$ for various lengths of the linear wire element $L_L$.}
\label{fig:current0}
\end{figure}

Figure~\ref{fig:current0} shows the ground-state current for the elementary device as a function of the magnetic flux $\Phi$ for a set of devices with various lengths of the linear part $L_L$ and fixed radius of the helix arc $R=500\,\mbox{nm}$.

The ground-state electric current~\eq{eq:current0} passing through the device depends linearly on the strength of the magnetic field $B$ at small values of the magnetic field, $B \ll B_0/(2\pi) \approx 10\,{\mathrm{G}}$:
\beqn
J_0 = {\tilde \sigma}_0 B \,, 
\label{eq:J0:B}
\eeqn
where the ground-state conductivity is 
\beqn
{\tilde \sigma}_0 & = & \frac{\pi e^2 L_R^4}{4 m L_H^2} \coth\frac{\pi L_L L_R}{2 L_H^2} \cdot \left[L_H^2 \sinh \frac{\pi L_L L_R}{L_H^2} \right.
\label{eq:sigma0}\\
& & \left. + \pi L_R \left(2 L_H \cosh^2\frac{\pi L_L L_R}{2 L_H^2} + L_L \right)\right]^{-1}\,.
\nonumber
\eeqn
Notice that Eq.~\eq{eq:J0:B} has the same form as the desired dissipationless transport law~\eq{eq:CME}.

\begin{figure}[!thb]
\begin{center}
\includegraphics[scale=0.125,clip=false]{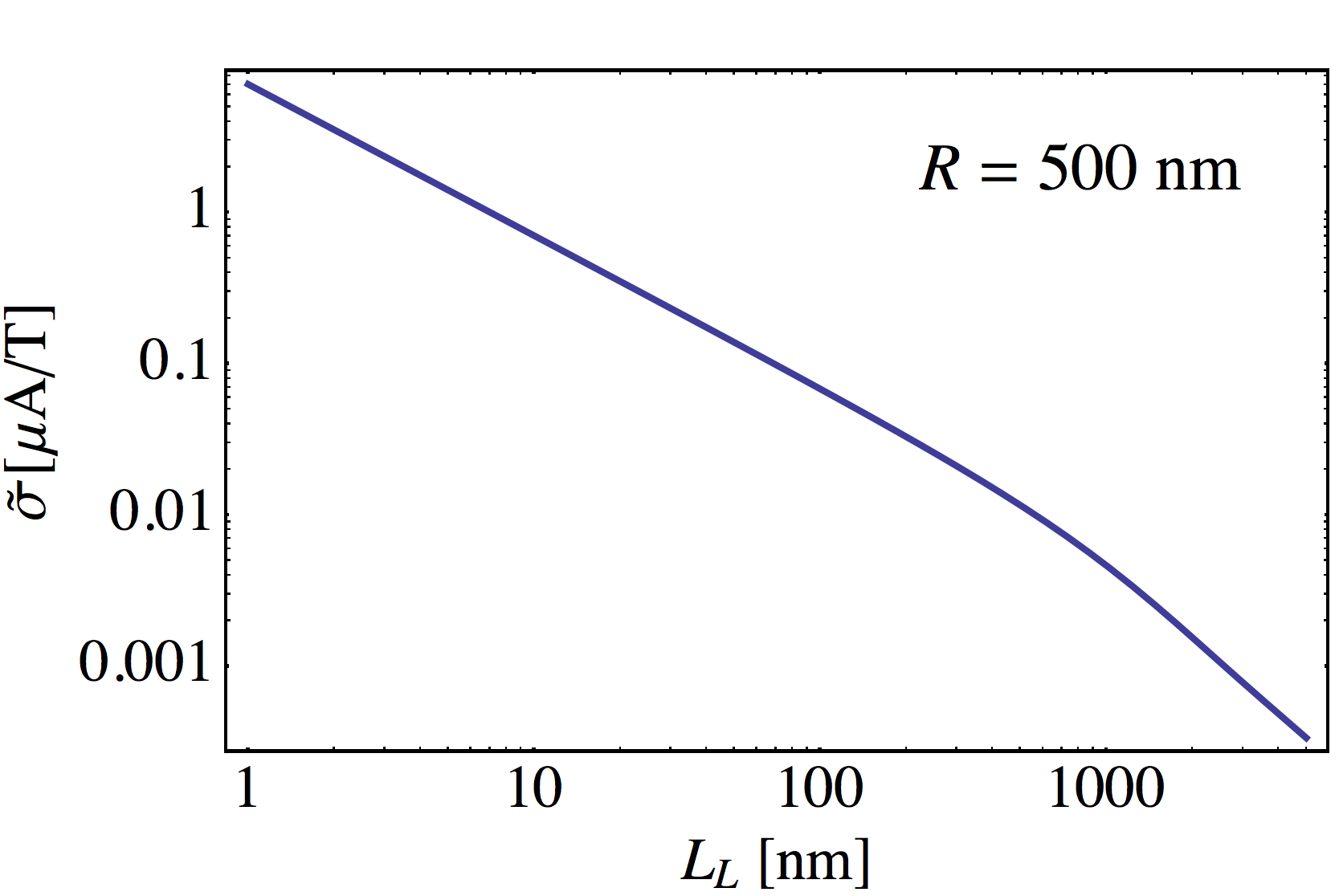} 
\end{center}
\vskip -5mm
\caption{The anomalous electrical conductivity $\tilde \sigma$ (in units of microAmpere per Tesla) for the one-wave ground-state solution, Eq.~\eq{eq:sigma0}, vs. the length of the linear wire element, $L_L$. The helical element has the fixed radius $R = 500$\,nm.}
\label{fig:conductivity}
\end{figure}

At certain specific quantized values of the magnetic flux, $\Phi = \Phi_0 l/2$, $l \in \Z$, the continuous energy spectrum is symmetric under the current reflections, $E(J) = E(-J)$, and the ground-state current is also zero. This feature is, in fact, anticipated from our experience with persistent currents in the normal metal nanorings. 

Summarizing this section, we would like to stress that there are two features of the spectrum of our elementary device (Fig.~\ref{fig:experiment}) which are different from those of the nanoring (Fig.~\ref{fig:quantumring}). In our device:
\begin{enumerate}
\item[(i)] the electric current flows through the device along the direction of the magnetic field (instead of being confined to the ring); 
\item[(ii)] the spectrum of the excitations is continuous (as compared to the discrete spectrum in the nanoring) so that the net equilibrium current passing through the device is always zero (see, however, the next section). 
\end{enumerate}

However, there are also two striking similarities in the qualitative features of the energy spectra of these two systems for generic values of the magnetic flux $\Phi$: 
\begin{enumerate}
\item[(a)] the inversion-asymmetric behavior single-level energies $E(J) \neq E(-J)$ and 
\item[(b)] the nonvanishing value of the electric current in the ground state $J(E_0) \neq 0$.
\end{enumerate}

Although properties (a) and (b) support the existence of the persistent equilibrium current through the device, the continuous nature of the spectrum, property (ii), leads to the vanishing net current. However, in the next section we show that for a long periodic structure made of the discussed short helical arcs, Fig.~\ref{fig:reconstruction}, the energy spectrum becomes discontinuous, so that the long device should support a persistent electric current along the direction of the magnetic field in accordance with anticipated dissipationless transport law~\eq{eq:CME}.

\section{Discrete asymmetric structure of energy levels in long wire}
\label{sec:long}

\subsection{Boundary conditions at junctions}

\begin{figure}[!thb]
\begin{center}
\includegraphics[scale=0.2,clip=false]{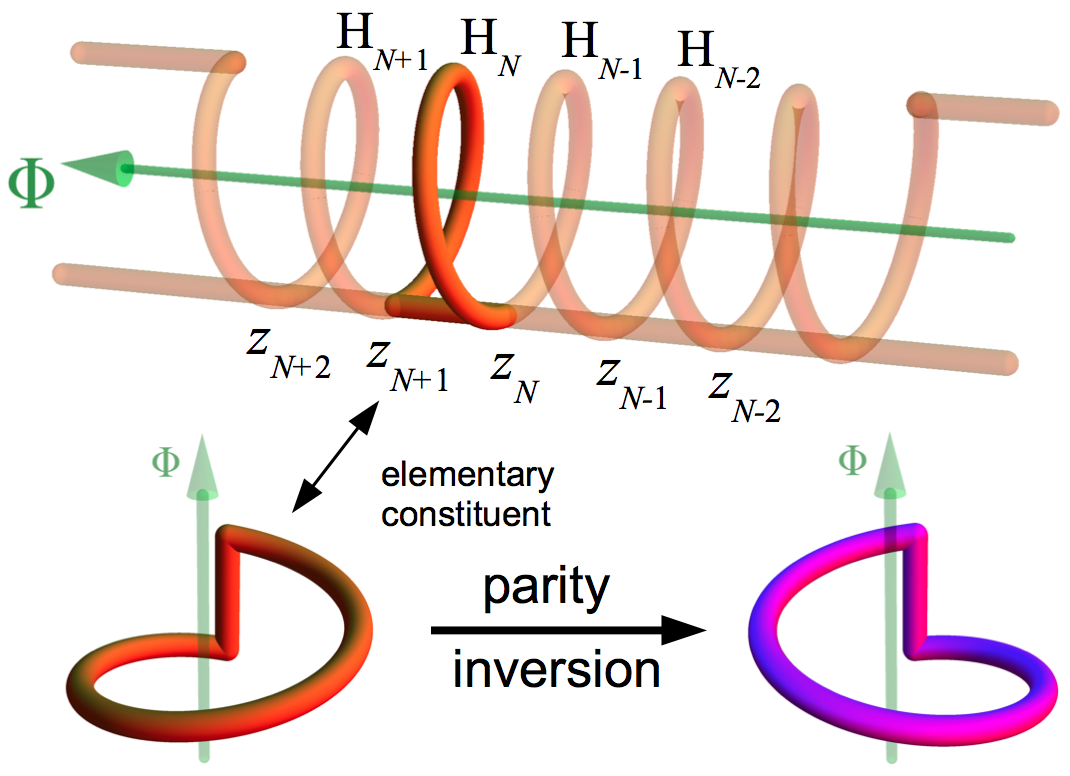} 
\end{center}
\vskip -5mm
\caption{Upper plot: illustration of a multidevice assembled from an (infinite) array of the elementary devices, Fig.~\ref{fig:experiment}. Lower plot: any elementary constituent of the multidevice (and the multidevice itself) is not invariant under parity inversion.}
\label{fig:reconstruction}
\end{figure}

Now let us consider the ``multidevice'', shown in Fig.~\ref{fig:reconstruction}. The multidevice is made of an infinite straight structure of the elementary devices (Fig.~\ref{fig:experiment}) $H_1\,, H_2\, \dots$ which are assembled together. The boundary conditions of the helix and straight-wire wave functions at the junction points $z = z_l \equiv L_L \, l$ and $\xi = \xi_l = L_H \, l$ are as follows:
\beqn
\Psi_H (L_H l) = \Psi_L (L_L l)\,, \qquad l \in \Z\,.
\label{eq:matching:multi}
\eeqn
These boundary conditions are also supplemented by the requirement that the wavefunctions and their derivatives of the helix and the line parts are smooth functions of the corresponding coordinates  $\xi$ and $z$ across each junction. 

The boundary conditions~\eq{eq:matching:multi} along with the smoothness conditions satisfy both standard Griffith\cite{ref:Griffith} and Shapiro\cite{ref:Shapiro} schemes for wavefunctions at the junctions. 

\begin{figure}[!thb]
\begin{center}
\includegraphics[scale=0.35,clip=false]{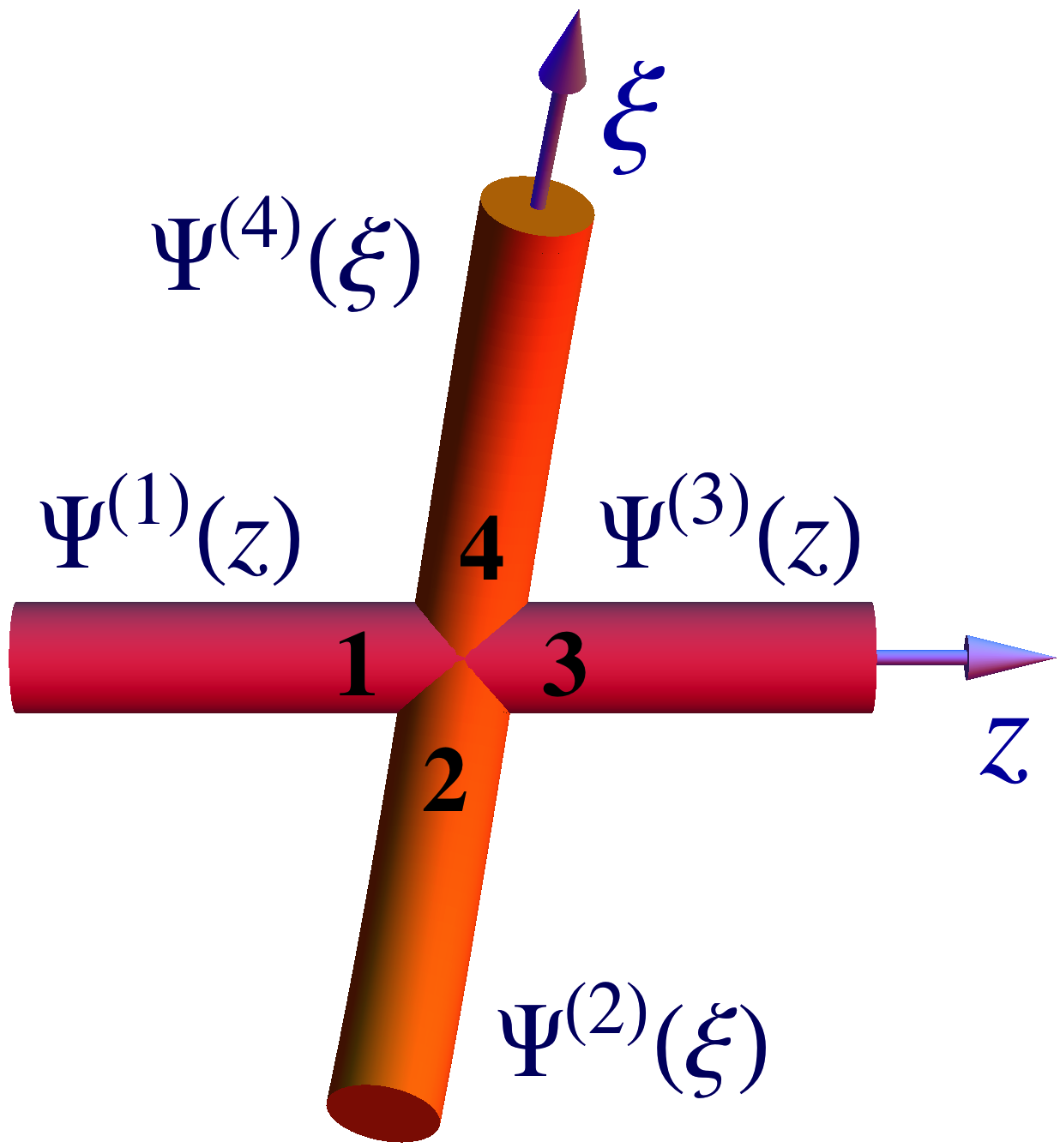} 
\end{center}
\vskip -5mm
\caption{The line (1 and 3) and helix (2 and 4) branches at the junction of the line and helix parts of the device.}
\label{fig:junction}
\end{figure}

The Griffith scheme imposes the unitarity condition that no net electric current flows into the junction\cite{ref:Griffith}:
\beqn
\sum_{i=1}^4 D_i \Psi^{(i)} = 0\,.
\label{eq:Griffith}
\eeqn
Here $D_i = \partial_i - i e A_i^\parallel$ is the covariant derivative along the $i$th branch, $A_i^\parallel$ is the component of the vector field parallel to the $i$th branch. The the labeling scheme for the branches is shown in Fig.~\ref{fig:junction}. Due to the mentioned smoothness conditions, the Griffith requirement~\eq{eq:Griffith} is explicitly satisfied by our conditions because of the conservation of the electric current separately in the line [$\Psi^{(1)}(z)$ and $\Psi^{(3)}(z)$] and helix [$\Psi^{(2)}(\xi)$ and $\Psi^{(4)}(\xi)$]  branches at the junction. 

The Shapiro scheme\cite{ref:Shapiro} requires the unitarity of the scattering matrix $S$ which relates the amplitudes for the ingoing waves
and the outgoing waves at the junction: $\psi_{\mathrm{out}}(z_l,\xi_l)  = S \psi_{\mathrm{in}}(z_l,\xi_l)$. Our boundary conditions imply that the amplitudes of the ingoing (outgoing) waves at the branches 1 and 2 are the same as the ones of the outgoing (ingoing) waves at the branches 3 and 4. Using the natural notation for the amplitudes, $\psi = (\Psi^{(1)}, \dots, \Psi^{(4)})^T$, the scattering matrix at each junction can be written in the following unitary ($S^\dagger S = \bbbone$) matrix form:
\beqn
S = 
\left(
\begin{array}{cccc}
0 & 0 & 1 & 0 \\
0 & 0 & 0 & 1 \\
1 & 0 & 0 & 0 \\
0 & 1 & 0 & 0 
\end{array}
\right)\,.
\label{eq:S}
\eeqn

\subsection{Solutions and spectrum}

Let us first solve the matching conditions for the one-wave anzats~\eq{eq:anzats:elementary} with $C^B_H=0$. In this case Eq.~\eq{eq:matching:multi} can be put in the following explicit form:
\beqn
C^A_H = C^A_L e^{i \alpha_q l} + C^B_L e^{i \beta_q l}\,,  \qquad l \in \Z\,,
\label{eq:multi:equation}
\eeqn
where
\beqn
\alpha_q & = & \gamma L_H - L_H q_H + L_L q_L\,, 
\label{eq:alpha:q}\\
\beta_q & = & \gamma L_H - L_H q_H - L_L q_L\,.
\label{eq:beta:q}
\eeqn
Setting $l = 0$  and $l=1$ in Eq.~\eq{eq:multi:equation} one gets, respectively, the relation:
\beqn
C^A_H & = & C^A_L + C^B_L\,,
\label{eq:AH:A:B}
\eeqn
supplemented with one of the following relations:
\beqn
C^B_L & = & \frac{1 - e^{i \alpha_q}}{e^{i \beta_q} - 1} C^A_L \qquad \mbox{if\, $\beta_q \neq 2 \pi n$\,,  \ $n \in \Z$}\,,
\label{eq:A:B:1}\\
{\mbox{or}} \quad & & \nonumber\\
C^A_L & = & \frac{e^{i \beta_q} - 1}{1 - e^{i \alpha_q}} C^B_L \qquad \mbox{if\, $\alpha_q \neq 2 \pi n$\,, \ $n \in \Z$}\,. \quad
\label{eq:A:B:2}
\eeqn
Then, using Eqs.~\eq{eq:multi:equation}, \eq{eq:AH:A:B}, \eq{eq:A:B:1} and \eq{eq:A:B:2} one gets the following condition:
\beqn
e^{i \beta_q} - e^{i \alpha_q} = \left( e^{i \beta_q} - 1 \right) e^{i \alpha_q l} + \left(1 - e^{i \alpha_q}\right) e^{i \beta_q l}, \quad
\label{eq:all:n}
\eeqn
which should be satisfied for all integer $l$.

There are three sets of solutions of Eq.~\eq{eq:all:n}:
\beqn
{\mathrm{(I)}} \quad \alpha_q & = & - 2 \pi n\,, \quad n \in \Z\,, \quad \beta_q \mbox{ is arbitrary},\ \ \\
{\mathrm{(II)}} \quad \beta_q & = & - 2 \pi n \,, \quad n \in \Z\,, \quad \alpha_q \mbox{ is arbitrary},\ \ \\
{\mathrm{(III)}} \quad \alpha_q & = & \beta_q + 2 \pi n\,.
\eeqn
These equations, together with Eqs.~\eq{eq:curvature}, \eq{eq:energy}, \eq{eq:alpha:q} and \eq{eq:beta:q}, give us the following solutions\cite{footnote1}:
\beqn
{\mathrm{(I)}} \quad \left\{
\begin{array}{rcl}
q^{({\mathrm{I}},n)}_{L} & = &  L_H^{-1} L_R^{-2} \Bigl[ L_L L_H (\gamma L_H + 2 \pi n) \\
& & + \sqrt{(\gamma L_H + 2 \pi n)^2 L_H^4 - \pi^2 L_R^4}\Bigr], \\[2mm]
q^{({\mathrm{I}},n)}_{H} & = & L_H^{-1} \left[\gamma L_H + 2 \pi n + L_L q^{({\mathrm{I}},n)}_{L} \right], \\[2mm]
C^A_H & = & C^A_L\,, \qquad C^B_H = C^B_L = 0\,, 
\end{array}
\right. \quad
\label{eq:q:solution:I}
\eeqn
\beqn
{\mathrm{(II)}} \quad \left\{
\begin{array}{rcl}
q^{({\mathrm{II}},n)}_{L} & = & L_H^{-1} L_R^{-2}  \Bigl[ - L_L L_H (\gamma L_H + 2 \pi n) \\
& & + \sqrt{(\gamma L_H + 2 \pi n)^2 L_H^4 - \pi^2 L_R^4}\Bigr], \\[2mm]
q^{({\mathrm{II}},n)}_{H} & = & L_H^{-1} \left[\gamma L_H + 2 \pi n - L_L q^{({\mathrm{II}},n)}_{L} \right], \\[2mm]
C^A_H & = & C^B_L\,, \qquad C^B_H = C^A_L = 0\,, 
\end{array}
\right. \quad
\label{eq:q:solution:II}
\eeqn
\beqn
{\mathrm{(III)}} \quad \left\{
\begin{array}{rcl}
q^{({\mathrm{III}},n)}_{L} & = & \frac{\pi n}{L_L}\,, \\[2mm]
C^A_H & = & C^B_H = 0\,, \qquad C^A_L + C^B_L = 0\,,
\end{array}
\right. \qquad \quad
\eeqn
and the corresponding values of the momentum in the helix arc $q_H$ are determined from Eqs.~\eq{eq:curvature} and \eq{eq:energy}:
\beqn
q_H^2 = q_L^2 + \frac{\pi^2 L_R^2}{L_H^4}\,.
\eeqn
The normalizability requirement imposes the following restriction on the values of the integer variable $n$ for solutions (I) and (II):
\beqn
L_H^2 \left| \gamma L_H + 2 \pi n\right| \equiv 2 \pi L_H^2 \left| \Phi/\Phi_0 + n\right| \geqslant \pi L^2_R\,.
\label{eq:constraint}
\eeqn

It is readily seen that the third solution $(III)$ corresponds to the standing waves in the straight wire part of our multidevice. There is yet another solution which corresponds to the standing waves in the helix arc part of the device:
\beqn
{\mathrm{(IV)}} \quad \left\{
\begin{array}{rcl}
q^{({\mathrm{IV}},n)}_{H} & = & \frac{\pi n}{L_H}\,, \\[2mm]
C^A_L & = & C^B_L = 0\,, \qquad C^A_H + C^B_H = 0\,.
\end{array}
\right. \qquad \quad
\eeqn

Using the normalization condition~\eq{eq:normalization}, we get the following explicit form of the solutions:
\beqn
\Psi^{\mathrm{(I)}}_{H,n}(\xi) & = & \frac{1}{\sqrt{L_H + L_L}} \exp\left\{i \left[q^{({\mathrm{I}},n)}_{H} - \gamma \right] \xi\right\}\,,
\label{eq:psi:H:I}\\
\Psi^{\mathrm{(I)}}_{L,n}(z) & = & \frac{1}{\sqrt{L_H + L_L}} \exp\left\{i q^{({\mathrm{I}},n)}_{L}  z \right\}\,,
\label{eq:psi:L:I}
\eeqn
\beqn
\Psi^{\mathrm{(II)}}_{H,n}(\xi) & = & \frac{1}{\sqrt{L_H + L_L}} \exp\left\{i \left[q^{({\mathrm{II}},n)}_{H} - \gamma \right] \xi\right\}\,,
\label{eq:psi:H:II}\\
\Psi^{\mathrm{(II)}}_{L,n}(z) & = & \frac{1}{\sqrt{L_H + L_L}} \exp\left\{- i q^{({\mathrm{II}},n)}_{L} z \right\}\,,
\label{eq:psi:L:II}
\eeqn
\beqn
\Psi^{\mathrm{(III)}}_{H,n}(\xi) & = & 0 \,,
\label{eq:psi:H:III}\\
\Psi^{\mathrm{(III)}}_{L,n}(z) & = & \sqrt{\frac{2}{L_L}} \sin \frac{\pi n z}{L_L} \,,
\label{eq:psi:L:III}
\eeqn
\beqn
\Psi^{\mathrm{(IV)}}_{H,n}(\xi) & = & \sqrt{\frac{2}{L_H}} e^{- i \gamma \xi} \sin \frac{\pi n \xi}{L_H} \,,
\label{eq:psi:H:IV}\\
\Psi^{\mathrm{(IV)}}_{L,n}(z) & = & 0 \,.
\label{eq:psi:L:IV}
\eeqn

The energies of these solutions are as follows ($n \in \Z$):
\beqn
E^{\mathrm{(I)}}_{n} & = & \frac{\hbar^2}{2 m} \left( q^{({\mathrm{I}},n)}_{L} \right)^2\,, 
\label{eq:E:I:n}\\
E^{\mathrm{(II)}}_{n} & = & \frac{\hbar^2}{2 m} \left( q^{({\mathrm{II}},n)}_{L} \right)^2\,, 
\label{eq:E:II:n}\\
E^{\mathrm{(III)}}_n & = & \frac{h^2 n^2}{8 m L_L^2}\,,
\label{eq:E:III:n}\\
E^{\mathrm{(IV)}}_n & = & \frac{h^2 n^2}{8 m L_H^2}\,,
\label{eq:E:IV:n}
\eeqn
where the momenta $q_L$ for solutions (I) and (II) are given in Eqs.~\eq{eq:q:solution:I} and \eq{eq:q:solution:II}, respectively.

Using Eqs.~\eq{eq:LH:L} and \eq{eq:gamma:Phi} we get the corresponding total single-level currents passing through the device for solutions (I) and (II):
\beqn
J^{\mathrm{(I),(II)}}_{n} & = & - \frac{e h}{m} \frac{1}{L_R^2} \left[\left( \frac{\Phi}{\Phi_0} + n \right) 
\right. \nonumber\\ & & \left. 
\pm \sqrt{\left( \frac{\Phi}{\Phi_0} + n \right)^2 - \frac{L_R^4}{4 L_H^4}}\right]\,,
\label{eq:J:I:tot:n} 
\eeqn
while the standing wave solutions give the vanishing total current:
\beqn
J^{\mathrm{(III)}}_{n} = J^{\mathrm{(IV)}}_{n} = 0\,.
\label{eq:J:IV:tot:n}
\eeqn
Due to the constraint~\eq{eq:constraint} the current~\eq{eq:J:I:tot:n} is always a real quantity.

Thus, the first solution, given in Eqs.~\eq{eq:psi:H:I} and \eq{eq:psi:L:I}, provides us with a generally nonzero electric current~\eq{eq:J:I:tot:n} passing through the device in the direction of the magnetic field. This longitudinal current is parametrically twice bigger than the transverse current circulating in the closed metallic ring of the same radius $R$, Eq.~\eq{eq:i:n}. All other solutions correspond to the vanishing total electric current.

\subsection{Persistent current in the periodic long wire}

We come to the important conclusion: the energy spectrum~\eq{eq:E:I:n} and the corresponding single-level currents~\eq{eq:J:I:tot:n} of the asymmetric (with respect to inversions $J \to -J$) modes~(I) in the multidevice are equivalent, up to unessential multiplicative factors, to the energies and single-level currents of the closed metallic ring, Eqs.~\eq{eq:E:n} and \eq{eq:i:n}, respectively. 

\begin{figure}[!thb]
\begin{center}
\includegraphics[scale=0.12,clip=false]{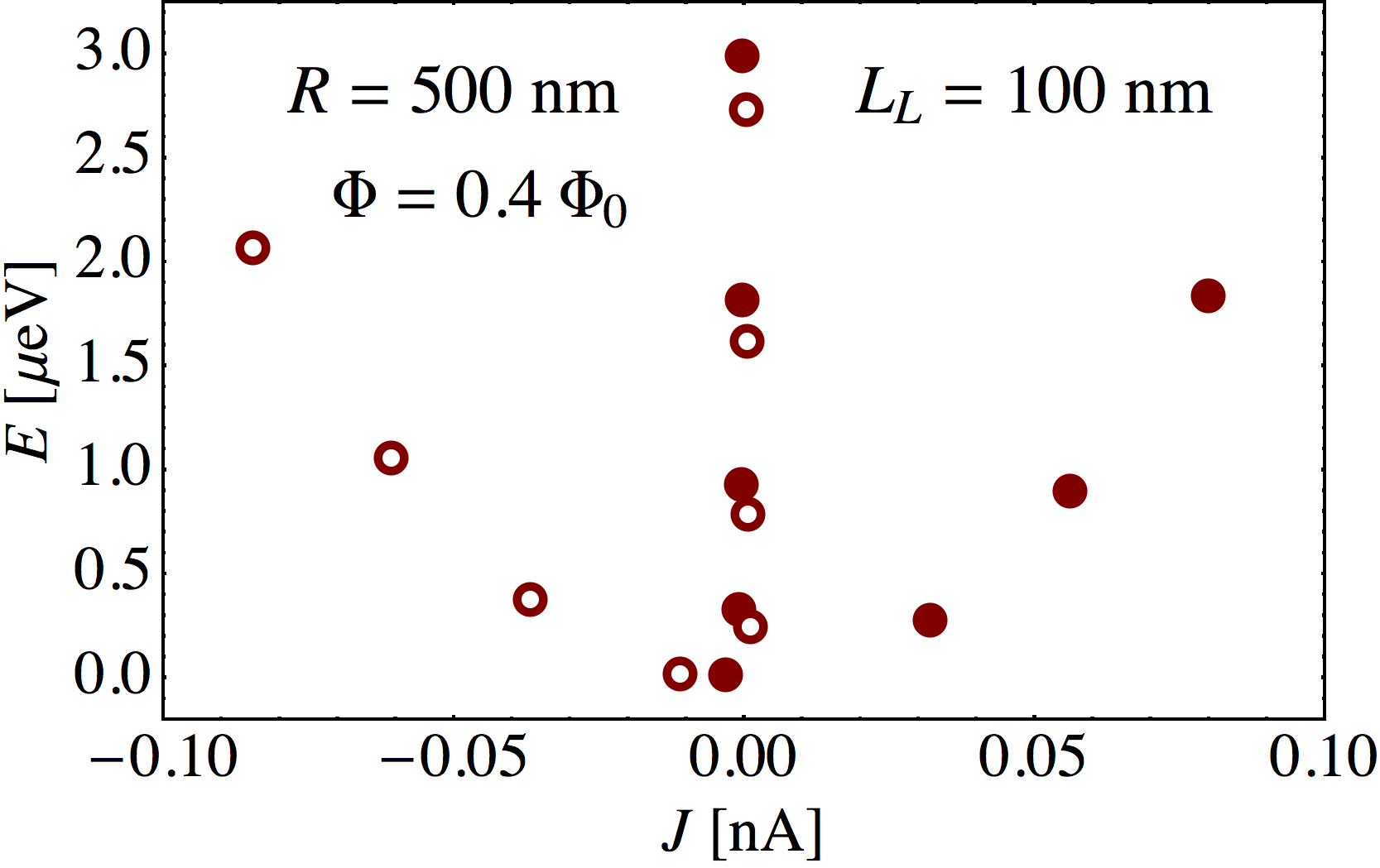} 
\end{center}
\caption{Energy of the elementary excitations vs. the corresponding single-level currents induced in the multidevice (with $R = 500\,\mbox{nm}$ and $L_L = 100\,\mbox{nm}$) by the magnetic flux $\Phi = 0.4\, \Phi_0$. The filled and open circles correspond to solutions~\eq{eq:q:solution:I} and \eq{eq:q:solution:II}, respectively. The zero-current states are not shown.}
\label{fig:multispectrum}
\end{figure}

It is the asymmetric modes which, as we expect, should provide us with a nonzero persistent current in our multidevice. However, the crucial difference between the nanoring setup (shown in Fig.~\ref{fig:quantumring}) and our multidevice (shown in Fig.~\ref{fig:device}) is that the currents in our proposal are transferred along the direction of the magnetic field instead of being confined to the ring itself. The presence of the standing modes (III) and (IV) is not essential for our qualitative discussion. 

Let us consider certain limits of our solutions (I) -- (IV). 

The ring-like limit $L_L \to 0$ is, in fact, a singular point of our solutions because the case $L_L = 0$ corresponds to a degenerate case of matching conditions~\eq{eq:matching:multi} (all helix arcs are transformed to rings which mutually overlap infinite number of times). Thus, the $L_L \to 0$ limit should not correspond to a system with $L_L = 0$. Nevertheless, in this limit the helix-arc solutions (I) and (II) converge to the anticipated periodic ring solution, so that their wavefunctions~\eq{eq:psi:H:I} and \eq{eq:psi:L:I}, energies~\eq{eq:E:I:n} and \eq{eq:E:II:n}, and the singe-level helix currents are converging to those of the ring solutions, Eqs.~\eq{eq:psi:n}, \eq{eq:E:n} and \eq{eq:i:n}, respectively. Notice that in the $L_L \to 0$ limit, the wavefunction of the line wire element is still nontrivial because, strictly speaking, our open system is not reduced completely in this singular limit to the closed isolated ring. 

The standing-wave solutions (III) and (IV), which carry no currents both in arc and line elements, should decouple from the spectrum because the energy of the former~\eq{eq:E:III:n} is divergent in the ring limit while the ring-limit of the latter becomes singular in our thin-wire approximation.

In the ``straight''  limit $L_L \to \infty$ one of the solutions (I) or (II), depending on the sign of the sum $\Phi/\Phi_0 + n$, decouples from the spectrum because its energy~\eq{eq:E:I:n} becomes infinitely large. The other solution reduces, after a proper renormalisation, to a plane-wave wavefunction shared between two straight parallel wires. Then the continuous wave vector $k$ is identified with $\pi l/l_L$ and the effect of the magnetic flux is negligible as the factor $\gamma \propto 1 / L_H$ vanishes in this limit. The standing-wave solutions (III) and (IV) survive the straight-wire limit as well. 

Finally, we would like to ask an important question: What are the symmetry arguments which may explain the very emergence of the dissipationless currents in the proposed devices? In the Introduction we have mentioned two specific examples of physical systems, where the dissipationless law~\eq{eq:CME} is expected to work: these are chirally-imbalanced quark-gluon plasma created in heavy-ion collisions and Weyl semi-metals. In both these cases the transport law~\eq{eq:CME} appears as a result of the non-invariance of the media with respect to the parity transformations. The parity odd nature of these systems is achieved at the microscopic level due to imbalance between left-handed and right-handed charge carries (quarks in heavy-ion collisions and charged fermionic excitations in Weyl semi-metals, respectively).

In our case, the parity-oddness of the proposed nanowire devices is achieved at the geometrical level. Both the elementary device and the multidevice are not invariant under spatial inversion, as it is illustrated in the lower panel of Fig.~\ref{fig:reconstruction}. Thus, the appearance of the dissipationless transport law~\eq{eq:CME} is also supported by the general symmetry arguments.

Finalizing this section we would like to mention that the energy--current spectra of the ring system, Fig.~\ref{fig:current:energy}, and the periodic wire system, Fig.~\ref{fig:multispectrum}, are very similar to each other: (i) they are both given by the parity-odd discrete quantities; (ii) in both cases the energy exhibits the parabolic dependence on the single-level current for large currents; and (iii) the physical scales of the energies and currents are the same. Thus, the persistent currents in both systems should be of the same order. We expect that the long periodic wire should exhibit the persistent current in the range from $1\,{\mathrm{pA}}$ to $1\,{\mathrm{nA}}$ for the magnetic fields up to a few Tesla scale.

\section{A chiral nanoribbon as a long planar dissipationless conductor}
\label{sec:twodim}

\begin{figure}[!thb]
\begin{center}
\includegraphics[scale=0.5,clip=false]{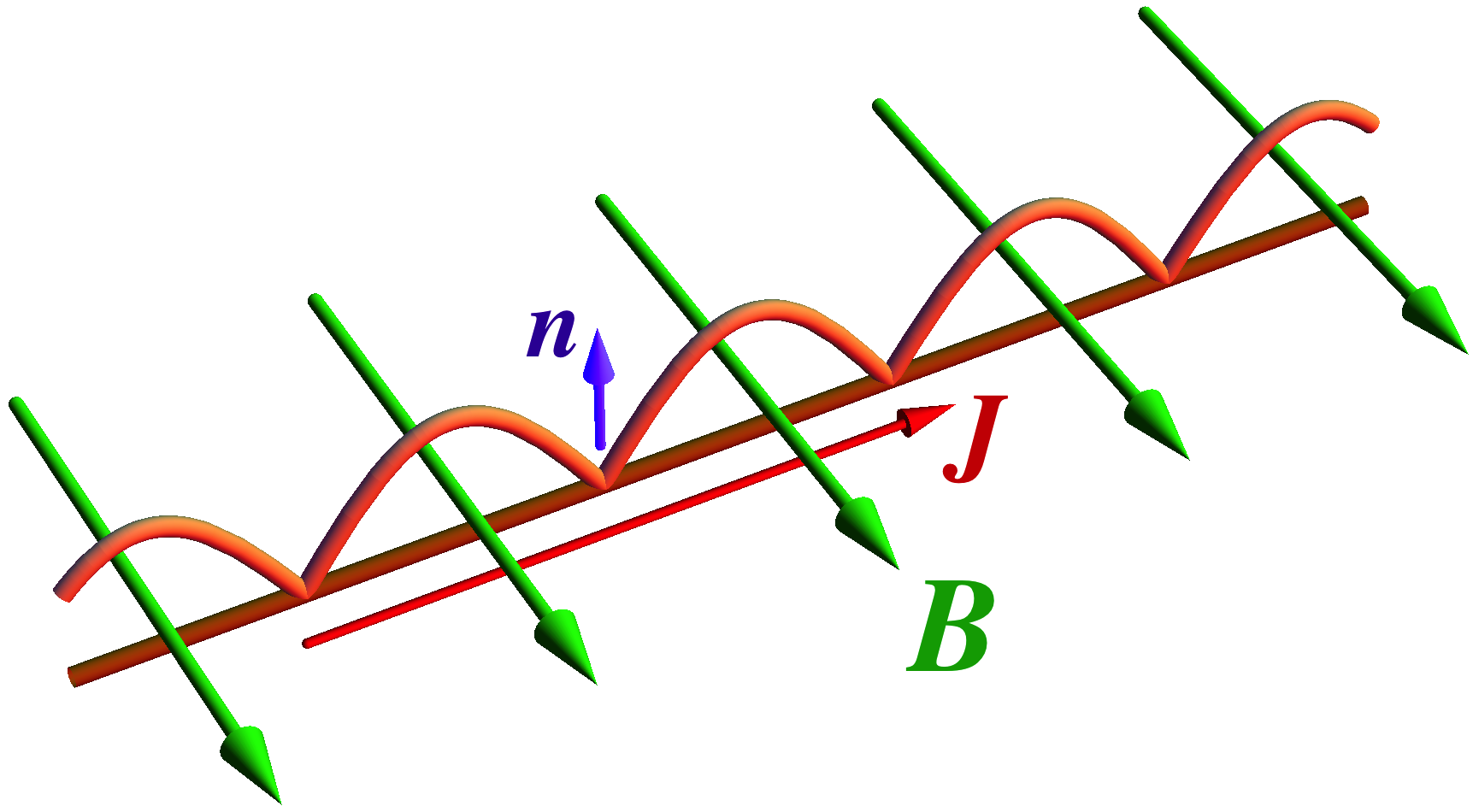} 
\end{center}
\caption{A chiral nanoribbon in the external magnetic field $\bs B$ may support a dissipationless electric current ${\bs J}$ according to Eq.~\eq{eq:alternative:CME}. The unit vector $\bs n$ characterizes the chirality direction of the nanoribbon.}
\label{fig:twodim}
\end{figure}

The three-dimensional multidevice, Figs.~\ref{fig:device}, has a two-dimensional (planar) counterpart which is illustrated in Fig.~\ref{fig:twodim}. In the planar structure, the arcs of the helix part of the three-dimensional wire structure are projected into a two-dimensional plane. The plane incorporates also the line part of the original wire structure. The magnetic flux is set to be normal to the surface of the plane in order to support a nonzero flux of the magnetic field through the areas bounded by the curved and linear wire elements\cite{footnote2}. The matching of the wavefunctions at the junctions is described by the continuity equation~\eq{eq:matching:multi} supplemented by the scattering matrix~\eq{eq:S}. 

The wire structure of Fig.~\ref{fig:twodim} can be treated a chiral nanoribbon because the parallel boundaries are not equivalent to each other. The difference in boundaries makes it possible to identify the constant vector $\bs n$ which marks the chirality direction of the nanoribbon.

The calculation of the energy-current spectrum of this system, Fig.~\ref{fig:twodim}, is identical to the one of the periodic wire shown in Fig.~\ref{fig:device}. Following the steps of derivation of the previous section we come to the conclusion that the energy spectrum of the electrons in the nanoribbon is a discrete parity-odd function of the electric current passing through the nanoribbon (in fact, the energy-current relation for the nanoribbon is given by Fig.~\ref{fig:multispectrum} for a suitable choice of the parameters $L_H$ and $L_L$). 

Thus, the nanoribbon of Fig.~\ref{fig:twodim} should support a dissipationless electric current in thermodynamic equilibrium. The corresponding transport law in the nanoribbon is given by the following alternative to Eq.~\eq{eq:CME}:
\beqn
{\bs J} = \sigma_\chi \,  {\bs n} \times {\bs B}\,.
\label{eq:alternative:CME}
\eeqn
This equation is invariant under time inversion, $t \to -t$ similarly to the London law~\eq{eq:London} in superconductor and to the anomalous transport law~\eq{eq:CME} discussed earlier in this paper. There is no entropy production associated with the electric current $\bs J$ so that this current is a dissipationless quantity.

In Eq.~\eq{eq:alternative:CME} the anomalous conductivity $\sigma_\chi$ is a parity-even quantity since both ${\bs J} $ and ${\bs n} \times {\bs B}$ behave as vectors under the spatial inversion, ${\boldsymbol x} \to - {\boldsymbol x}$.

\section{Discussion and Conclusions}
\label{sec:conclusions}

In this article we have proposed two long nanostructures, shown schematically in Figs.~\ref{fig:device} and \ref{fig:twodim}, which should support dissipationless equilibrium electric current. The persistent electric current in these devices -- made of resistive metallic nanowires -- is supposed to be supported by applied magnetic field similarly to the persistent electric current in normal metal nanorings. According to Eqs.~\eq{eq:CME} or \eq{eq:alternative:CME}, the electric current should be proportional to the strength of the magnetic field if the magnetic field is low enough. As the magnetic field becomes stronger, the electric current should experience periodic oscillations. The devices are not invariant under the inversion of parity (see, for example, Fig.~\ref{fig:reconstruction}), so that the parity-odd nature of these systems is achieved at the geometrical level. 

It is worth stressing that the parity-odd energy spectrum of the proposed nanowire structures has the discrete nature despite these structures are macroscopically large systems. It is the discontinuities of the spectrum which make possible the existence of the equilibrium dissipationless current. 

Thus, we have shown that the equilibrium, magnetic-field-induced persistent current is protected by both topological and symmetry-related arguments. We have shown that the energy levels and single-level electric currents in our devices share the same properties (and physical scales) as the energy levels and currents in the nanorings.

In our article we have discussed the very basic features of the idealized devices. As we may expect from our experience with the normal metal nanorings, the suggested dissipationless currents in the real devices may be suppressed by various corrections emerging due to finite width of the wire, thermal effects, and effects of disorder and interactions. However, following our both theoretical and experimental experience gained with the metallic nanorings, we think that these effects should not destroy the dissipationless current completely.

In addition, we would like to mention that different but nominally identical nanorings exhibit different persistent currents (both in magnitudes and signs) because the persistent current in nanorings is, basically, a few electron phenomenon~\cite{ref:Science,ref:PRL}. Contrary to the nanorings, our devices are open systems so that the few-electron properties may be less important. However, we expect that the thermal diffusive suppression (caused by impurities) can still make a substantial quantitative contribution to the anomalous conductivities $\tilde \sigma$ and $\sigma_\chi$ of our devices.

One may suggest that the proposed devices can also be made of metallic carbon nanotubes as the persistent currents were also suggested to occur in carbon nanotori~\cite{ref:carbon:nanotori}.

Given the success of the modern experiments with normal metal nanorings, the fabrication of the proposed devices is within the reach of current technology. We hope that the dissipationless equilibrium electric current may be observed experimentally. We expect that the persistent current should be in the range from $1\,{\mathrm{pA}}$ to $1\,{\mathrm{nA}}$ for the magnetic fields up to a few Tesla scale.

\acknowledgments 
The author is grateful to Karel Van Acoleyen for interesting discussions, to Karl Landsteiner and to Alexey Nikulov for making the author aware of Refs.~[\onlinecite{ref:Weyl}] and [\onlinecite{ref:Alexey}], respectively, and to Marcel Franz for correspondence. The work was supported by Grant No. ANR-10-JCJC-0408 HYPERMAG.


\begin{thebibliography}{99}
\bibitem{footnote0} There are systems which do not possess a discrete spectrum while being able to support a persistent current due to the presence of a finite number of discontinuities in otherwise continuous energy spectrum. A relevant example can be found in Ref.~[\onlinecite{ref:Kibis}].
\bibitem{footnote1}
In fact, these equations give, formally, a larger set of solutions, which, however, are equivalent to solutions~\eq{eq:q:solution:I} and~\eq{eq:q:solution:II}.
\bibitem{footnote2}
From the point of view of the discrete reflection symmetries, our proposal of the dissipationless current in the nanoribbon of Fig.~\ref{fig:twodim} has a qualitative similarity with a chain-like flat system of asymmetric Aharonov-Bohm rings studied in Ref.~[\onlinecite{ref:Alexey}].
\end{thebibliography}
\end{document}